\documentclass[reprint, amsmath,amssymb, aps,prb, nofootinbib, longbibliography]{revtex4-2}

\usepackage{graphicx}
\usepackage{dcolumn}
\usepackage{natbib}
\usepackage{bm}
\usepackage{braket}
\usepackage[colorlinks=true,linktoc=page,linkcolor=blue,citecolor=blue,urlcolor=blue]{hyperref}
\usepackage{amsthm}
\usepackage{xcolor}
\usepackage{bbm}
\usepackage{soul}
\usepackage{cleveref}
\let\oldref\ref
\renewcommand{\ref}[1]{\pare{\oldref{#1}}}
\usepackage{amsfonts}

\begin{document}

\title{Mixed higher-order topology and nodal and nodeless flat band topological phases in a superconducting multiorbital model}

%\title{First-, second-, and hybrid-order topologies and nodal and nodeless flat band topological phases in a superconducting multiorbital model}

\author{Rodrigo Arouca$^1$}
\author{Tanay Nag$^{2,1}$}\email{tanay.nag@hyderabad.bits-pilani.ac.in}
\author{Annica M. Black-Schaffer$^1$}
\affiliation{$^1$Department of Physics and Astronomy, Uppsala University, SE 75120 Uppsala, Sweden}
\affiliation{$^2$Department of Physics, BITS Pilani-Hyderabad Campus, Telangana 500078, India}

\date{\today}

\begin{abstract}
    We investigate the topological phases that appear in an orbital version of the Benalcazar-Bernevig-Hughes (BBH) model in the presence of conventional spin-singlet $s$-wave superconductivity and with the possibility of tuning an in-plane magnetic field. We chart out the phase diagram by considering different boundary conditions, with the topology of the individual phases further examined by considering both the Wannier and entanglement spectra, as well as the Majorana polarization. 
    For weak to moderate values of magnetic field and superconducting pairing amplitude, we find a second-order topological superconductor phase with eight zero-energy corner modes. Further increasing field or pairing, half of the corner states can be turned into zero-energy edge-localized modes, thus forming what we name hybrid-order phase. Then, we find two different putative first-order topological phases, a nodal and a nodeless phase, both with zero-energy flat bands localized along mirror-symmetric open edges. For the nodal phase, the flat bands are, as expected, localized between the nodes in reciprocal space, while in the nodeless phase, the zero-energy boundary flat band instead spans the whole Brillouin zone and appears disjoint from the fully gapped bulk spectrum. As a consequence, this model present several unexpected phases with unusual surface states that can be tuned via an external magnetic field.
\end{abstract}
\maketitle

%%%%%%%%%%%%%%%%%%%%%%%%%%%
% SECTION: INTRODUCTION
%%%%%%%%%%%%%%%%%%%%%%%%%%%
\section{Introduction}\label{sec_intro}

The study of topological materials is an extremely active area of research in condensed matter physics. They present phases of matter that are not characterized by spontaneous symmetry breaking but rather by topological invariants. In the Altland-Zirnbauer classification \cite{altland1997nonstandard, schnyder2008classification}, time-reversal, particle-hole, and chiral symmetries classify ten possible topological classes with bulk energy gaps, indicating the kind of invariant and the branches of the symmetry-protected boundary states. The number of possible symmetry-protected topological classes of free fermions has further been increased by including crystalline symmetries to this original classification \cite{fu2011topological, slager2013space, PhysRevX.7.041069,po2017symmetry, neupert2018topological}. 

In addition to new topological classes, crystalline symmetries also allow for the presence of higher-order topological phases \cite{benalcazar17_PRB, benalcazar2017_Science, schindler2018higheradv, schindler2018higher}, where the topological invariant computed in the bulk is not related to modes appearing on the whole surface of the material but rather on a smaller set. As an example, the original model of a higher-order topological insulator, the two-dimensional (2D) Benalcazar-Benervig-Hughes (BBH) model \cite{benalcazar17_PRB, benalcazar2017_Science}, hosts a second-order topological phase with protected zero-energy modes appearing at the corners of the system. The last years have seen a profusion of work on higher-order topological insulators \cite{ezawa2018higher,PhysRevB.98.081110,PhysRevB.99.245151, PhysRevResearch.1.032045,PhysRevB.101.161116,PhysRevResearch.2.012067, arouca2020thermodynamics,Saha_2023,Nag21,Ghosh22b,Dumitru19, PhysRevLett.130.116204}, with experimental realization in materials \cite{schindler2018higher, noguchi2021evidence} and a variety of metamaterials \cite{imhof2018topolectrical,serra2019observation, ni2019observation,xue2019acoustic, el2019corner,kempkes2019robust, cerjan2020observation, zheng2022observation,li2022higher, Experiment3DHOTI.aSonicCrystals, Schulz2022}. 

A particularly interesting class of higher-order topological systems is higher-order topological superconductors (HOTSC), both for static and driven Hamiltonians \cite{wang2018weak, PhysRevB.106.L060307, Wang23edgecorner,Ghosh21,Ghosh21b,Ghosh22,Ghosh22c,Roy20}. The corner or hinge states in HOTSC appear at zero energy and can, as such, be Majorana zero modes (MZMs) \cite{kitaev2001unpaired,lutchyn2010majorana, oreg2010helical, mourik2012signatures,nadj2013proposal,nadj2014observation,nadj2013proposal, Baldo_2023, PhysRevB.107.184519} since they present an equal amount of particle and hole components and can thus  their own antiparticles, promising for applications in quantum computing \cite{RevModPhys.80.1083}. This however requires tuning the degree of degeneracy for the zero-energy states, as well as controlling their spatial extent and separation. 

Topologically protected boundary states can also appear in nodal superconductors \cite{Lofwander_2001, PhysRevLett.89.077002,  PhysRevB.85.024522,matsuura2013protected, PhysRevLett.112.117002,schnyder2015topological,Debmalya2022, bonetti2023van, PhysRevB.97.174522,doi:10.1143/JPSJ.81.011013,PhysRevB.83.224511}. Topological nodal superconductivity is a non-trivial phase that is however not contained in the Altland-Zirnbauer classification since the bulk of the system is gapless at the nodal points \cite{PhysRevB.85.024522,matsuura2013protected,schnyder2015topological}. Nevertheless, the presence of the nodal points is protected by symmetry, and there also exists a bulk-boundary correspondence between the topology of the bulk nodal points and boundary-localized flat bands located between the nodes. 

It would be interesting to uncover systems where multiple different superconducting topological phases are readily realized, including higher-order topology and various nodal states. This would both provide  realizations of different individual topological phases and possible intriguing combinations thereof, possibly even uncovering previously unknown phases, and also offer tunability between the different phases and their characteristic properties. To be precise, being motivated by both recent studies on the BBH model in the normal, i.e.~non-superconducting, state and already existing HOTSC \cite{Ghosh21, Wu22, Yan18,maiellaro2021topological}, we in this work seek the uncover the different superconducting phases in the BBH model in the presence of a tunable magnetic field. We are concerned with mapping out the full phase diagram, and, in particular, we focus on unexpected topological superconducting phases generated by the intricate interplay between the higher-order topology of the normal state and superconductivity and magnetic field.

\begin{figure}[!t]
    \includegraphics[width=\linewidth]{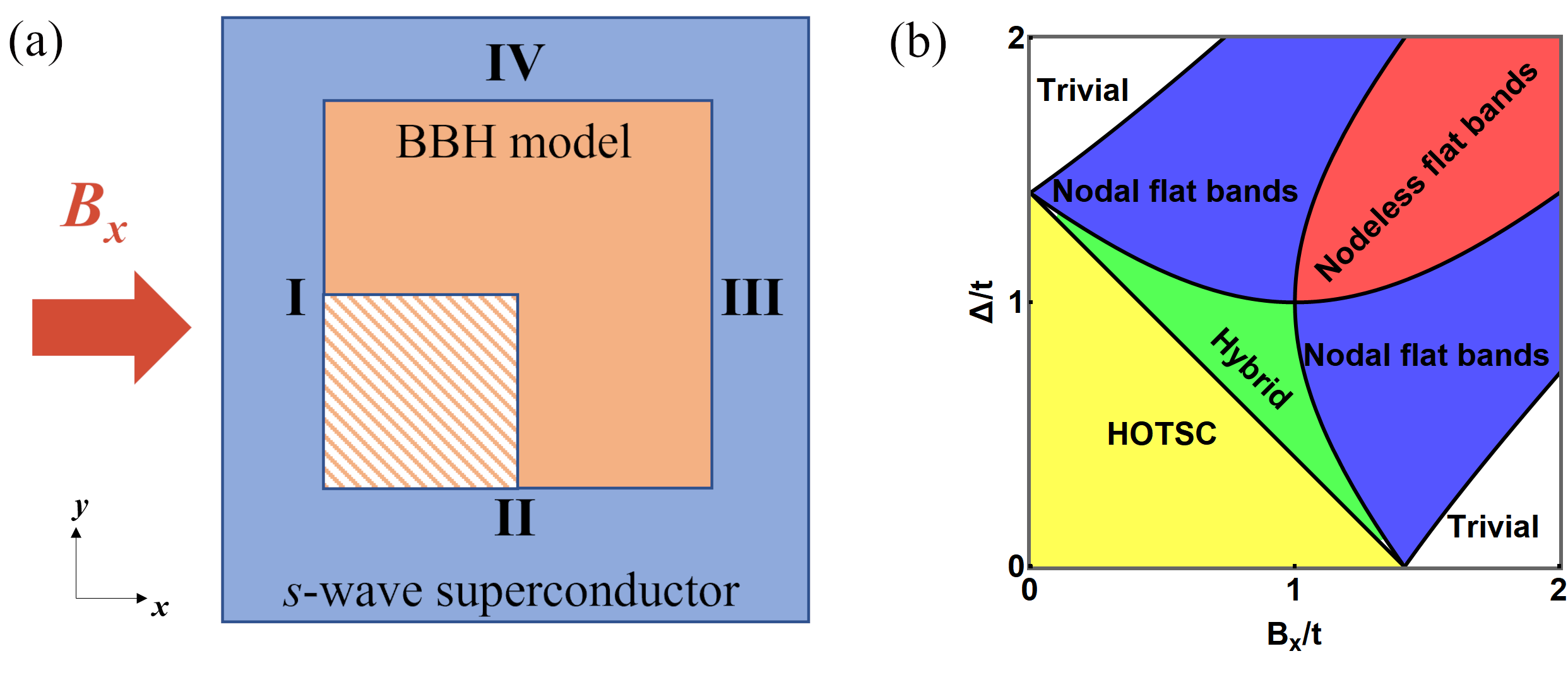}
    \caption{(a) Schematic realization of the superconducting BBH model: BBH system with conventional spin-singlet $s$-wave superconducting pairing $\Delta$ induced by proximity effect from a substrate and an in-plane magnetic field $B_x$. Different edges are indicated by Roman numerals. Dashed area represents the region where the entanglement spectrum is computed. (b) Summary of phase diagram: Solid lines represent analytical expressions for the phase boundaries, while different colors represent different topological phases. See main text for further definitions.}
    \label{fig_phase}
\end{figure}

Aiming at least for higher-order topology, we choose to investigate the orbital version of the BBH model in the deep topological limit with conventional spin-singlet $s$-wave superconductivity induced by proximity effect from a substrate and using an applied in-plane magnetic field as an additional easily accessible tunable parameter, all illustrated in Fig.~\ref{fig_phase}(a). 
The in-plane magnetic field $B_x$ breaks the $C_4$ symmetry responsible for protecting the corner states in the BBH model, while the proximity-induced superconducting order parameter $\Delta$ transforms these states in electron-hole excitations. Moreover, we find that the superconducting term in this orbital model is represented by an unusual matrix structure, giving rise to a multitude of different topological phases. Using the Wannier spectrum \cite{neupert2018topological, benalcazar2017_Science, schindler2018higher}, the entanglement spectrum \cite{hui08ent, hughes2011inversion, Fukui18entpol,you20hotent, Zhu20Cnent} for a quarter of the lattice indicated as in Fig.~\ref{fig_phase}(a), and the Majorana polarization \cite{Sticlet12majpol, bena2017testing}, we completely characterize the topological phases and obtain the rich phase diagram in Fig.~\ref{fig_phase}(b). 

To briefly summarize the phase diagram, for $B_x=0$ and $\Delta \approx 0$ the result is a superconducting version of the second-order topological phase of the BBH model, characterized by eight corner states and displayed as the HOTSC phase (yellow) in Fig.~\ref{fig_phase}(b). The pairing makes the corner states of the BBH model transform into Andreev bound states, built up from two degenerate MZMs, located at each corner. 
The presence of both finite pairing and an in-plane magnetic field makes even more interesting and unexpected topological phases appear. For larger values of both $\Delta$ and $B_x$ we first find another higher-order topological phase, a type of hybrid ordered phase (green). In this phase, some (four) of the original zero-energy corner states stay, becoming isolated MZMs, while the rest turn into MZMs localized along on the edges in the $y$-direction, edges II and IV, in Fig.~\ref{fig_phase}(a). This presents an intriguing mix, or hybrid, of a second-order and a dipolar topological phase, but where the number of edge localized states do not grow with system size, as usually expected. Further increasing $B_x$ and $\Delta$, we find two other also interesting topological phases. These are both first-order topological phases with flat bands at zero energy as their boundary modes. One of this phases is a traditional nodal phase: a nodal flat band phase (blue), where the bulk of the system is gapless with symmetry-protected nodes and the zero-energy flat bands, localized on the $y$-edge, occur in the region of momentum space between the nodes, just as expected for a topological nodal superconductor. In contrast, in the other topological flat band phase the bulk stays completely gapped, but we nevertheless still find flat bands at zero energy, now spanning across the whole Brillouin zone and localized at the edges along the $x$-direction. We name this latter, unexpected, phase a nodeless flat band phase (red). The association of these flat bands to a bulk invariant remains unclear, but we find that they clearly appear in a quantized Wannier spectrum. Both flat band phases present four zero-energy states per momenta, two at each edge, with a large Majorana polarization, such that we consider them to be MZMs. 

Our results show how a seemingly simple model, a normal state with spin-orbit coupling together with proximity-induced conventional superconductivity and in-plane magnetic field, can generate a plethora of different topological phases. The different kind of surface states can further be accessible by tuning physical parameters, in particular the magnetic field that can be externally controlled. Importantly, our results establish the existence of both a hybrid-order phase and a fully gapped (i.e.~nodeless) topological phase with zero-energy flat bands boundary states, in addition to already known topological phases. Our findings thus increase the catalogue of emergent topological superconducting phases and establish their microscopic origin. 

The rest of this work is structured as follows. In Sec.~\ref{sec_model} we briefly summarize the main properties of the BBH model and discuss the form of the Hamiltonian, especially the superconducting pairing and the magnetic field term, which we add to the BBH model. In Sec.~\ref{sec_topo}, we introduce the three topological indicators we use in this work, explaining what aspects of topological phases can be characterized by each invariant. In Sec.~\ref{sec_phases} we report our main results. First, we consider gap closings as a function of $\Delta$ and $B_x$, which indicate the phase boundaries. For each phase, we analyze the behavior of the energy spectra for different boundary conditions as well as the different topological invariants, in order to characterize all phases. In Sec.~\ref{sec_disc} we discuss the dual \cite{sato09dual} and surface \cite{Ghosh21, bonetti2023van} Hamiltonians for this model, which help us understand the presence of nodes starting with $s$-wave pairing and the higher-order phases of the model. Finally, in Sec.~\ref{sec_conc} we conclude and summarize our main results. 

%%%%%%%%%%%%%%%%%%%%%%%%%%%
% SECTION: MODEL
%%%%%%%%%%%%%%%%%%%%%%%%%%%
\section{Model}\label{sec_model}

The paradigmatic model of a higher-order topological insulator, the BBH model \cite{benalcazar2017_Science, benalcazar17_PRB}, is described by the Bloch Hamiltonian
\begin{equation}
    h_{\rm BBH}(\mathbf{k})=\mathbf{d}(\mathbf{k})\cdot \bm {\gamma},
    \label{eq_h_BBH}
\end{equation}
where $d_1=t \sin(k_y a)$, $d_2=\lambda+t \cos(k_y a)$, $d_3=t \sin(k_x a)$, $d_4=\lambda+t\cos(k_x a)$, $\gamma_i=\sigma_2\otimes s_i$ for $i=1, 2, 3$,  and $\gamma_4=\sigma_1\otimes s_0$. The model is defined on a square lattice with a lattice parameter $a$. In addition to chiral ($\Pi=\sigma_3\otimes s_0$) symmetry\footnote{We remark that the original model BBH model is also time-reversal and particle-hole symmetric. However, since we are dealing with spinfull fermions instead of spinless ones, the representations of time-reversal and particle-hole symmetry are different, and these symmetries are broken by one of the terms in the model.}, this model is also reflection symmetric along $x$ ($M_x= i \sigma_1\otimes s_3$) and $y$ ($M_y= i \sigma_1\otimes s_1$), besides presenting $C_2$ ($r_2=M_x M_y=-i \, \sigma_0 \otimes s_2$) and $C_4$ \footnote{This is the case when the hoppings along $x$ are the same as the hopping along $y$. When the hoppings are different, only $C_2$ symmetry is present \cite{benalcazar17_PRB}.} ($r_4=(\sigma_1+i \sigma_2)/2\otimes s_0-i (\sigma_1-i \sigma_2)/2\otimes s_2$) symmetries \cite{benalcazar17_PRB}. For $\lambda<t$, this model presents topological corner modes protected by both chiral and $C_4$ symmetries.

The original BBH model can be considered as the collection of spinless fermions on a square lattice with four sublattices, being an extension of the two-dimensional (2D) Su-Schrieffer-Heeger model \cite{PhysRevLett.42.1698} with a $\pi$-flux.  Here, we instead interpret $\sigma_i$ and $s_i$ as Pauli matrices ($\sigma_0=s_0=\mathbbm{1}_2$) associated with orbital and spin degree of freedom, respectively. This Hamiltonian then has spin-orbit coupling between orbital and spin on the same site (terms proportional to $\lambda$) and between different sites (terms proportional to $t$). 
This choice is motivated when considering the proximity effect from a conventional superconductor, where the pairing is $s$-wave (onsite) intra-orbital (proportional to $\sigma_0$) spin-singlet (proportional to $s_y$), which is not possible for spinless fermions. We here further consider the deep topological limit of Eq.~\eqref{eq_h_BBH}, by setting $\lambda=0$. This both generates nontrivial topology in the normal state and removes the on-site spin-orbit coupling.

Since it is known that a conventional superconductor with spin-orbit coupling can host topological superconductivity in an in-plane magnetic field \cite{sato09dual, lutchyn2010majorana,oreg2010helical}, we also add an in-plane external magnetic field $B_x$ along $x$, accounted for by a Zeeman term proportional to $s_x$. An interesting question then is to investigate the interplay between superconductivity and magnetic field with the intrinsic higher-order topology of the BBH model. The resulting Bogoliubov-de Gennes (BdG) Hamiltonian for this system in the particle-hole basis becomes
\begin{equation}
    h_{\rm BdG}(\mathbf{k})=\left(\begin{matrix}\mathbf{h}(\mathbf{k})&-i\mathbf{\Delta}\\i\mathbf{\Delta}&-\mathbf{h}^*(-\mathbf{k})\end{matrix}\right)=\, \mathbf{D}(\mathbf{k})\cdot \mathbf{\Gamma},
    \label{eq_h_BdG}
\end{equation}
where $h(\mathbf{k})=h_{\rm BBH}(\mathbf{k})+B_x \, \sigma_0\otimes s_1 $ is the total normal-state Hamiltonian. For the second equality we use the vectors $\mathbf{D}=(d_1, d_2, d_3, d_4, \Delta, B_x)$ and $\mathbf{\Gamma}$ to write the Hamiltonian in terms of matrices in the particle-hole, orbital, and spin degrees of freedom. Using $\tau_i$ as the Pauli matrices in the particle-hole degrees of freedom, we construct the matrices $\Gamma_i=\tau_3\otimes \gamma_i$, $\Gamma_4=\tau_3\otimes \gamma_4$, $\Gamma_5=\tau_2\otimes\sigma_0\otimes s_2$, and $\Gamma_6=\tau_3\otimes\sigma_0\otimes s_1$. Note that by this definition, $\mathbf{\Delta}=\Delta \sigma_0\otimes s_2$ which represents an intra-orbital spin-singlet pairing, as expected by proximity effect from an external conventional superconductor. This model has chiral $\Pi=\tau_1\otimes\sigma_0\otimes s_0$ and particle-hole symmetry $P=\tau_1\otimes \sigma_0\otimes s_0 K$. The discussion about crystalline symmetries is present in Sect.~\ref{subsec_wann}.

We remark that in the deep topological limit of the BBH model, which is the parameter range for $\lambda/t$ that we focus on in our work, only the corners are gapless in the normal state. Therefore, in a more realistic model, in which superconductivity is introduced self-consistently in the topological BBH system, pairing terms would technically only be present around the corners (or edges) of the system. However, since our main focus in this work is to understand the variety of topological phases that arise due to the interplay of the BBH model with superconductivity, we choose to add $s$-wave pairing without self-consistent treatment. Our approach follows similar recent treatments in the literature of higher-order topological superconductors, see, for instance, Refs.~\cite{Yan18, Wu22, Ghosh21}.

Before investigating the different phases of this system in detail, we next discuss the topological invariants that we use to characterize them.

%%%%%%%%%%%%%%%%%%%%%%%%%%%
% SECTION: TOPO INVARIANTS
%%%%%%%%%%%%%%%%%%%%%%%%%%%
\section{Topological characterization}\label{sec_topo}

As already hinted by the phase diagram in Fig.~\ref{fig_phase}(b), the model in Eq.~\eqref{eq_h_BdG} presents a variety of topological phases with a mix of first- and second-order, as well as nodal topological phases. In addition, since we are dealing with a spinful topological superconductor, we can have different symmetry-protected boundary states. Therefore, we do a thorough analysis of the topology using three different invariants, which can identify different aspects of the topological phases. 

We first use the Wannier spectrum \cite{benalcazar17_PRB, neupert2018topological} to investigate the presence of a non-trivial polarization in the lattice. Since the presence of higher-order topological phases is not completely characterized by the Wannier spectra, we also use a real-space indicator, the entanglement spectrum \cite{hui08ent, zhu2020identifying}, to verify whether the boundary modes are of higher-order topological origin. Finally, to understand whether these boundary states may be MZMs or ABS, we use the Majorana polarization \cite{Sticlet12majpol, bena2017testing}, which indicates how much of a combination of electron and hole a state in a superconductor is. In addition to these indicators, we study the energy spectra using different boundary conditions and the local density of states (LDOS) at zero energy to further characterize and especially verify the various phases and the phase transitions inbetween them. Below, we briefly review these tools, all already used in different previous settings and models, in order to provide a comprehensive background to be able to explain the signatures of each topological phase.

\subsection{Wannier spectrum}\label{subsec_wann}

In this subsection, to make this work more self-contained, we pedagogically review the use of the Wannier spectrum to characterize first- and higher-order topological phases as discussed, for instance, in Refs.~\cite{PhysRevB.47.1651, spaldin2012beginner, neupert2018topological}. In the modern theory of polarization, the charge polarization is directly related to the Berry phase \cite{PhysRevB.47.1651,PhysRevLett.80.1800, spaldin2012beginner, neupert2018topological} being, therefore, a topological property of a material. However, the direct numerical calculation of these quantities is not so practical due to an overall ill-definition of the phase of wavefunctions \cite{neupert2018topological}. A very convenient alternative is based on the use of Wilson loops \cite{PhysRevD.10.2445, benalcazar17_PRB, neupert2018topological}. For a 2D system with periodic boundary conditions along $x$ and open boundary conditions along $y$, the Wilson loop components are defined by \cite{neupert2018topological} \footnote{The Wilson loop, in general, needs to be defined in terms of a reference point $\mathbf{k}_i$ \cite{benalcazar17_PRB} from which the loop is made. However, for the topological invariants considered here, the Wannier spectrum is the same for all reference points, and we thus choose $k_i=-\pi$ for both $k_x$ and $k_y$.} 
\begin{equation}
   \left(W\right)_{mn}^{x}=\braket{u_m(\pi)|\prod\limits_{k_x}^{\pi \leftarrow -\pi}\mathcal{P}(k_x)|u_n(-\pi)}, 
   \label{eq_Wilson}
\end{equation}
where $\mathcal{P}$ is a projector
\begin{equation}
    \mathcal{P}(k_x)=\sum\limits_{m}\ket{u_m(k_x)}\bra{u_m(k_x)}
\end{equation} 
over the occupied eigenstates $\ket{u_m(k_x)}$ of the (semi-periodic) Hamiltonian with momentum $k_x$.

$W^{x}$ is a unitary matrix which can be associated with the so-called Wannier Hamiltonian $H_W^x$ 
\begin{equation}
   W^{x}=e^{i 2\pi H_{W}^x}.
   \label{eq_Wannier_Ham}
\end{equation}
The eigenvalues $\nu^x$ of $H_{W}^x$ are the Wannier spectrum of the system with open boundaries along $y$. A gauge transformation, associated with the change of phases of the wavefunctions, can change $\nu^x$ by integer values, making these quantities, in general, defined mod 1. We choose a gauge where $\nu^x$ take values between $0$ and $1$ \cite{neupert2018topological, benalcazar17_PRB}. The presence of crystalline symmetries imposes some constraints in the Wannier spectrum \cite{neupert2018topological, benalcazar17_PRB}. For instance, symmetry upon reflection along the $x$-axis makes $\nu^x$ come in $(\nu, 1-\nu)$ pairs  \cite{benalcazar17_PRB}. In this way, $\nu^x=0.5$ are reflection invariant and indicate the presence of boundary modes protected by this symmetry, created by a non-trivial polarization along $x$\cite{neupert2018topological, benalcazar17_PRB}. The same holds, \textit{mutatis mutandis}, to periodic boundary conditions along $y$ with boundaries open along $x$, obtaining a corresponding $\nu^y$. Thus, when either $\nu^x$ or $\nu^y$ presents modes at $0.5$, or half-quantized modes, we obtain a dipolar phase. For a higher-order topological phase, there are instead half-quantized midgap states in both $\nu^x$ and $\nu^y$ \cite{benalcazar17_PRB}. 

An illustrative example is the BBH model in Eq.~\eqref{eq_h_BBH}. This model presents reflection symmetry along both $x$ and $y$, which restricts $\nu^{x/y}$ to appear in pairs. For $\left|\lambda/t\right|<1$, the system is in a second-order, or quadrupolar, topological phase characterized by four corner modes at zero energy, corresponding four $0.5$ eigenvalues in both $\nu^x$ and $\nu^y$ \cite{benalcazar17_PRB}. For reference, for $\left|\lambda/t\right|>1$, the model is in a trivial phase, with no midgap states in both the energy and Wannier spectra \cite{benalcazar17_PRB}. If we further allow $t$ or $\lambda$ to be different along $x$ and $y$, we can obtain a phase with polarization just along one of the directions showing first-order topology. 

Before moving on, we discuss why we are not using the nested Wilson loop spectrum \cite{benalcazar17_PRB} to characterize the higher-order topological phases. The nested Wilson loop is computed by using the eigenvectors of the Wannier Hamiltonian $H_W$, Eq.~\eqref{eq_Wannier_Ham}, but for fully periodic boundary conditions in the general expression of the Wilson loop in Eq.~\eqref{eq_Wilson}. It is often taken as a clear indicator of higher-order topology, presenting midgap states in its spectrum when there are symmetry-protected higher-order modes \cite{benalcazar17_PRB}. For example, in the BBH model, Eq.~\eqref{eq_h_BBH}, the nested Wilson loop is a phase, which is equal to zero in the trivial phase and $\pi$ in the quadrupolar phase \cite{benalcazar17_PRB}. However, for the nested Wilson loop spectrum to present quantized values, one needs inversion symmetry. But our full system in Eq.~\eqref{eq_h_BdG} breaks mirror symmetry along $x$: $M_x h_{\rm BdG}(k_x,k_y)(M_x)^{-1} \ne h_{\rm BdG}(-k_x,k_y)$ with $M_x=\tau_3 \otimes \sigma_1 \otimes s_3$, while it preserves mirror symmetry along $y$: $M_y h_{\rm BdG}(k_x,k_y)(M_y)^{-1} = h_{\rm BdG}(k_x,-k_y)$ with $M_y=\tau_3 \otimes \sigma_1 \otimes s_1$. Consequently, inversion symmetry, generated by $\mathcal{I}=M_x M_y$, is broken as $\mathcal{I} h_{\rm BdG}(k_x,k_y) \mathcal{I}^{-1} \ne h_{\rm BdG}(-k_x,-k_y)$. Therefore, we cannot use the nested Wilson loop as a bulk invariant to diagnose our topological phases. In fact, we computed the nested Wilson loop for this model, which presents non-quantized values for any finite $\Delta$, reinforcing that it is not a good invariant. We instead revert to the Wannier spectra along $x$ and $y$.               

\subsection{Entanglement spectrum}\label{subsec_ent}

Since we cannot use the nested Wilson loop spectrum, but only the Wannier spectrum along $x$ or $y$, an alternative tool to characterize higher-order topological phases is useful\footnote{We remark that, in principle, a real space invariant related to a generalized chiral symmetry could also be used to characterize the topological phases in our model \cite{PhysRevLett.125.166801,PhysRevLett.128.127601,PhysRevB.107.075413}.}. One such tool has recently turned out to be the entanglement spectrum \cite{hui08ent, hughes11inv, Fukui18entpol, herviou2019entanglement, you20hotent, Zhu20Cnent, Ortega-Taberner22entnon}. In the same way that a non-trivial polarization in the lattice can be determined using the Wannier spectrum, it can also be diagnozed by the entanglement spectrum \cite{hui08ent, hughes11inv, Fukui18entpol, herviou2019entanglement, you20hotent, Zhu20Cnent, Ortega-Taberner22entnon}. To obtain the entanglement spectrum, we compute the correlation matrix in the occupied state $\ket{\Omega}$
\begin{equation}
    C_{\mathbf{r}, \tau, \sigma, s; \mathbf{r}', \tau', \sigma', s'}=\braket{\Omega|c^\dagger_{\mathbf{r}, \tau, \sigma, s}c_{\mathbf{r}', \tau', \sigma', s'}|\Omega},
\end{equation}
where $\ket{\Omega}$ represents the (many-body) fermionic ground state and $c^\dagger_{\mathbf{r}, \tau, \sigma, s}$ creates a particle ($\tau=1$) or hole ($\tau=-1$) in orbital $\sigma$ with spin $s$ at position $\mathbf{r}=(x,y)$. The entanglement spectrum $\xi$ consists of the eigenstates of the correlation function constrained to some finite region in real space. One can intuitively understand the cut(s) needed to create such as finite region as creating artificial boundaries in the system, such that the presence of boundary states may appear in the properties of the entanglement spectrum. For instance, for a system with inversion symmetry, the entanglement spectrum of a lattice cut in half displays modes at $0.5$ in the topological phase \cite{hughes2011inversion}, analogously to what happens to the Wannier spectrum.  A cut that preserves the symmetries that protect the corner modes can also be used to diagnose the presence of a higher-order topological phase \cite{you20hotent, zhu2020identifying}. Since these modes are protected by $C_4$ symmetry, we cut the system in half along both $x$ and $y$, obtaining a quarter of the original lattice, as indicated by the dashed area in Fig.~\ref{fig_phase}(a).

\subsection{Majorana polarization}\label{subsec_Maj}
To determine the Majorana nature of the boundary states the Majorana polarization $P$ is useful, defined as \cite{Sticlet12majpol, bena2017testing, PhysRevB.107.L201405}
\begin{equation}
    P_m(x,y)=\sum\limits_{\sigma, s}2 \psi_{x,y, \tau=1, \sigma, s; m} \psi_{x,y, \tau=-1, \sigma, s; m},
    \label{eq_def_P}
\end{equation}
where $\psi_{\mathbf{r},\tau=1, \sigma, s; m}$ ($\psi_{\mathbf{r}, \tau=-1, \sigma, s; m}$) is the particle (hole) component of the wavefunction of the  $m$-th eigenstate at position $\mathbf{r}=(x,y)$ with orbital $\sigma$ and spin $s$. This is a tool to characterize how much particle-hole symmetric a state is. In particular, the quantity  
\begin{equation}
    C_m=\frac{\left|\sum\limits_{x,y} P_m(x, y)\right|}{\sum\limits_{x,y, \tau, \sigma, s}|\psi_{x, y, \tau, \sigma, s;m}|^2},
    \label{eq_def_C}
\end{equation}
compares the value $P$ with the usual probability density of a state $m$. Therefore, $C$ quantifies how much of the wavefunction is particle-hole symmetric \cite{bena2017testing}. 

For systems that present only one isolated state per boundary, the Majorana polarization becomes an unambiguous indicator of a MZM. We note that, however, if there are many putative MZMs per boundary, since they are degenerate in energy, one may obtain different values of $P$ for different linear combinations of the states at zero energy. 
Thus, even if we numerically find a high value of the Majorana polarization compared to the probability density, it can still be unclear whether two such putative MZMs can actually recombine into a complex fermion. For such recombination to be able to not occur, different spin and orbital degrees of freedom generally have to be in play.
Such ambiguousness is the case for some topological phases in our system and we can thus not use $P$ as a completely unique indicator in these cases. Nevertheless, we still investigate the midgap states in terms of the Majorana polarization to provide an additional tool whenever it is distinctive. Since we always obtain a real $P$, we use only its real value to check its sign across the lattice. 

\subsection{Spectral characterization}\label{subsec_spec}

In addition to the topological invariants and indicators discussed above, it is useful to consider how the energy spectrum and the wavefunctions of the system behave in every phase. Symmetry-protected topological states normally appear at zero energy at the boundaries for a gapped or nodal bulk. Since, in our case, we have surface modes localized both on the edges and corners of the lattice, we extract the energy spectra for several different boundary conditions. For the bulk spectrum we apply fully periodic boundary conditions and generally sample the Brillouin zone taking paths connecting the high-symmetry points of the square lattice $\Gamma=(0,0)$, $X=(\pi/a, 0)$, $Y=(0, \pi/a)$, and $M=(\pi/a, \pi/a)$. For phases with edge or corner states we also apply open boundary conditions in the appropriate directions. For all midgap states we also plot the sum of the Majorana polarization for states at zero energy and compute $C$ in Eq.~\eqref{eq_def_C} to verify whether these states are MZMs or not. To obtain complete information on the localization of all low-lying states in the system, we also show the local density of states (LDOS) at zero energy or frequency $\omega=0$.

%%%%%%%%%%%%%%%%%%%%%%%%%%%
% SECTION: TOPO PHASES
%%%%%%%%%%%%%%%%%%%%%%%%%%%
\section{Topological phases}\label{sec_phases}

In this section, we present our main analysis of the topological phases of the model in Eq.~\eqref{eq_h_BdG} using the topological invariants discussed in Sec.\ref{sec_topo}. We focus on the deep topological limit of the BBH model, setting $\lambda=0$ for simplicity, which allow us to obtain analytical expressions of the phase boundaries, but we remark that the topological phases are present for $|\lambda/t|<1$. We refer to Appendix~\ref{app_phase_finite_lambda} for results on finite $\lambda$. For $\lambda=0$, the topological phase diagram is displayed in Fig.~\ref{fig_phase}(b). Here, we start by detailing how we obtain the phase boundaries, followed by a detailed description of each of the phases.

\begin{figure}[htb]
    \includegraphics[width=\linewidth]{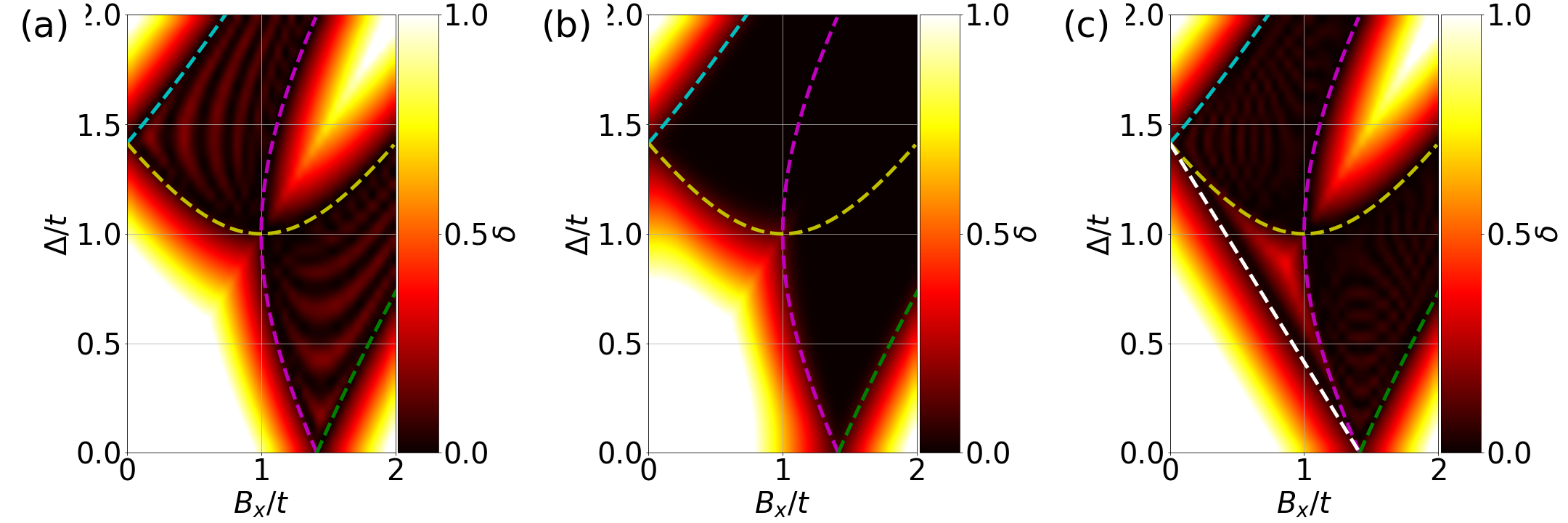}
    \caption{Energy gap $\delta$ between the highest valence band and lowest conduction band around zero energy, calculated from Eq.~\eqref{eq_h_BdG}, as a function of superconducting pair amplitude $\Delta/t$ and in-plane magnetic field $B_x/t$ for (a) fully-periodic, (b) $x$-periodic (open in $y$) and (c) $y$-periodic (open in $x$) boundary conditions. 
    Parameters used: $\lambda=0$, system size $32$ unit cells in each direction. Dashed lines represent analytical results, see main text for a description.}
    \label{fig_phase_PBC}
\end{figure}

\subsection{Phase boundaries}

A quantum phase transition is accompanied by the closing of the gap of the system \cite{vojta2003quantum}. Therefore, the topological phase boundaries are obtained by considering the energy gap, $\delta$ between the highest valence band and the lowest conduction band around zero energy. In Fig.~\ref{fig_phase_PBC} we plot $\delta$ as a function of the superconducting order parameter $\Delta$ and magnetic field $B_x$ for different boundary conditions. For fully periodic boundary conditions, we can additionally obtain the critical lines analytically using the Bloch Hamiltonian, which we represent in dashed lines in Fig.~\ref{fig_phase_PBC}(a), overlayed on the value of $\delta$ for each $\Delta$ and $B$ numerically computed using a real space Hamiltonian. For $B_x=\sqrt{2t^2+\Delta(\Delta-2t)}$ (magenta dashed line) the bulk spectrum closes at momenta $(k_x,k_y)=(0, \pi/2)$ and $(\pi, \pi/2)$ and for $\Delta=\sqrt{2t^2+B_x(B_x-2t)}$ (yellow dashed line) it closes at momentum $(\pi/2, \pi/2)$. These two lines separates the nodal flat bands phase (blue region in Fig.~\ref{fig_phase}(b)) from the nodeless flat bands phase (red region in Fig.~\ref{fig_phase}(b)) and hybrid phase. Further, the bulk spectrum also closes at $B_x=\sqrt{2t^2+\Delta(\Delta+2t)}$ (green dashed line) at momentum $(k_x,k_y)=(\pi, \pi/2)$ and at $\Delta=\sqrt{2t^2+B_x(B_x+2t)}$ (cyan dashed line) at momentum $(\pi/2, \pi/2)$, which separates the trivial phase from the nodal flat band phase. These results also establish that the HOTSC (yellow in Fig.~\ref{fig_phase}(b)), hybrid (green), trivial (white), and nodeless flat band (red) phases are all fully gapped in the bulk.

\begin{figure*}[htb]
    \includegraphics[width=\linewidth]{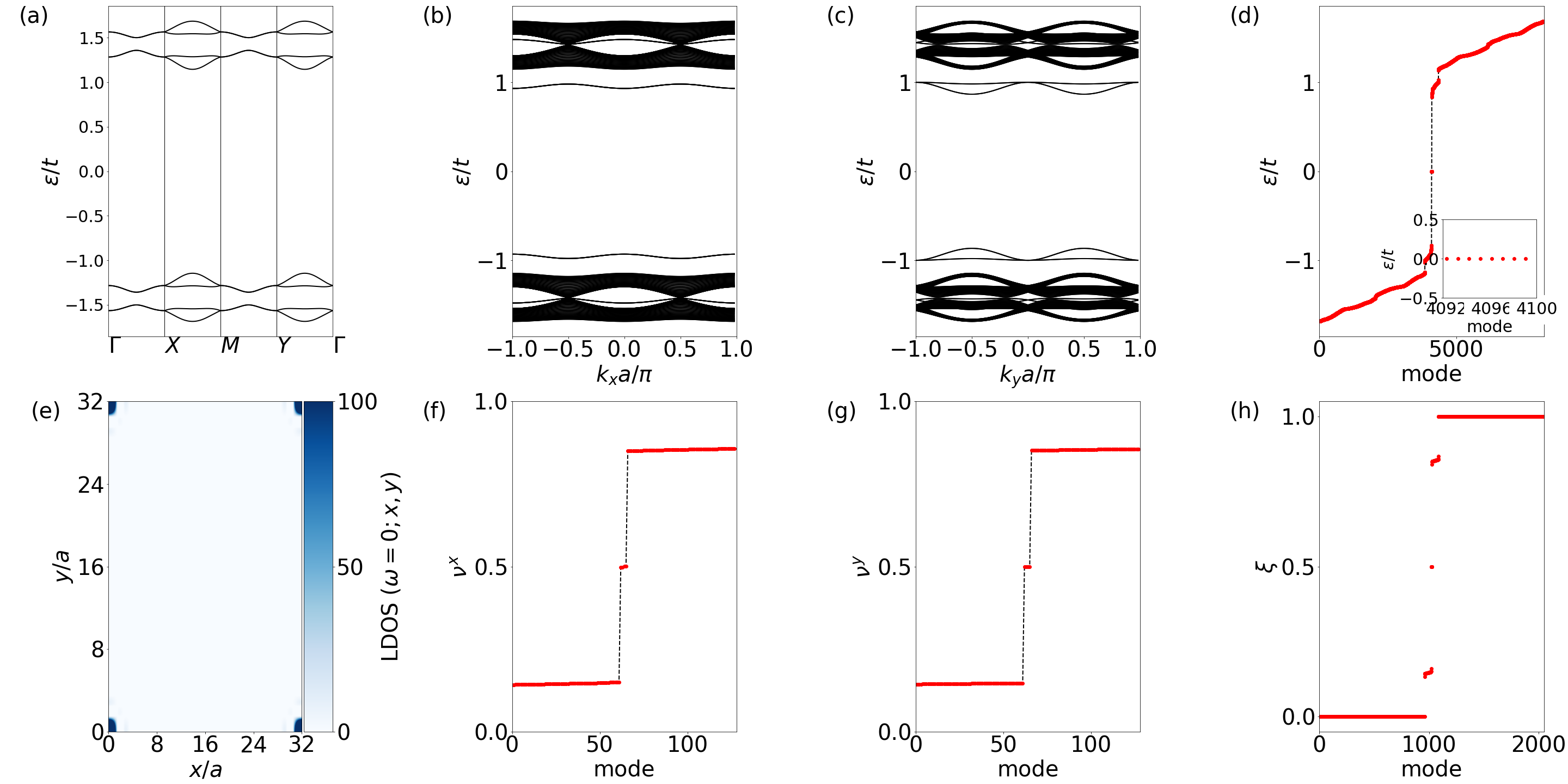}
    \caption{Main features of the HOTSC phase. Energy spectrum $\epsilon$ for fully periodic (a), $x$-periodic (b), $y$-periodic (c), and open (d) boundary conditions. Inset in (d) a zoom-in on the modes in the middle of the spectrum with energy $\epsilon=0$. LDOS at zero energy (e). Wannier spectrum along $x$ (f) and $y$ (g). Entanglement spectrum $\xi$ (h). Parameters used: $B_x=0.2 t$, $\Delta=0.1 t$, $\lambda=0$, system size $32$ unit cells for directions with open boundary conditions and $100$ $k-$points.}
    \label{fig_HOTSC_I}
\end{figure*}

To complement the results for fully periodic boundary conditions, we also analyze how the gap closes for open boundary conditions along $y$ in Fig.~\ref{fig_phase_PBC}(b)  and along $x$ in Fig.~\ref{fig_phase_PBC}(c). This analysis of different boundary conditions brings three important additional pieces of information.
First, we notice that there is a new gap-closing line at $\Delta=\sqrt{2}t-B_x$ (white dashed line), which indicates the phase boundary between the HOTSC and hybrid phases in Fig.~\ref{fig_phase}(b). This line just appears for $x$-open boundary conditions, see Fig.~\ref{fig_phase_PBC}(c), illustrating how the gap only closes along the $y$-direction at this phase transition. 
Second, we notice that the nodal flat bands phase (blue in Fig~\ref{fig_phase}(b)) seemingly hosts a small but finite gap, illustrated by the faint red arc-like features in Figs.~\ref{fig_phase_PBC}(a,c), but it does become completely gapless in Fig.~\ref{fig_phase_PBC}(b). An analysis of the spectrum of this system for semi-periodic boundary conditions in Sect.~\ref{subsec_nodal_FBM} show that, in fact, for all values of $B_x$ and $\Delta$ in this phase, the gap closes at $k_y=\pi/2$ and different $k_x$. Therefore, the spectrum is actually gapless in all three figures, and the red arcs are just finite-size effects in Figs.~\ref{fig_phase_PBC}(a,c). This verifies that the nodal flat bands phase is a bulk nodal phase.
Finally, we find that one of the bulk gapped regions, the nodeless flat band phase (red in Fig.~\ref{fig_phase}(b)) is also gapped for $x$-open boundary conditions but notably not for $y$-open boundary conditions, as seen in Fig.~\ref{fig_phase_PBC}(b). This indicates zero-energy states localized to edges along the $x$-direction.

In the next subsections, we detail the properties of each of the non-trivial topological phases, while the trivial phase is discussed in Appendix~\ref{app_triv}. We characterize the general properties considering both the energy spectra extracted above and the topological invariants discussed in Sec.~\ref{sec_topo}, for representative values of $B_x$ and $\Delta$ in all phases. 
\begin{figure}[htb]
    \includegraphics[width=0.75\linewidth]{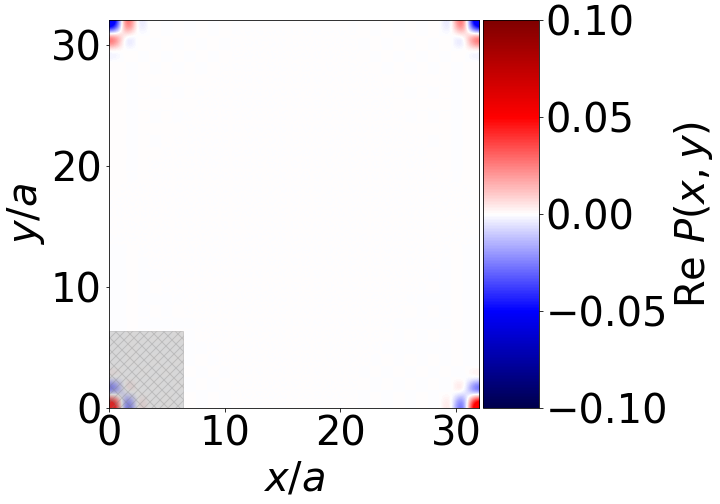}
    \caption{Sum of the Majorana polarizations $P$ of the HOTSC phase boundary states. The gray dashed area indicates the region where $C$ is calculated. For the corner states localized in this corner, $C$ is small (see text for discussion), indicating that these states are not MZMs. Parameters same as in Fig.~\ref{fig_HOTSC_I}.}
    \label{fig_HOTSC_I_modes}
\end{figure}

\subsection{HOTSC phase}\label{subsec_HOTSC_I}

\begin{figure*}[htb]
    \includegraphics[width=\linewidth]{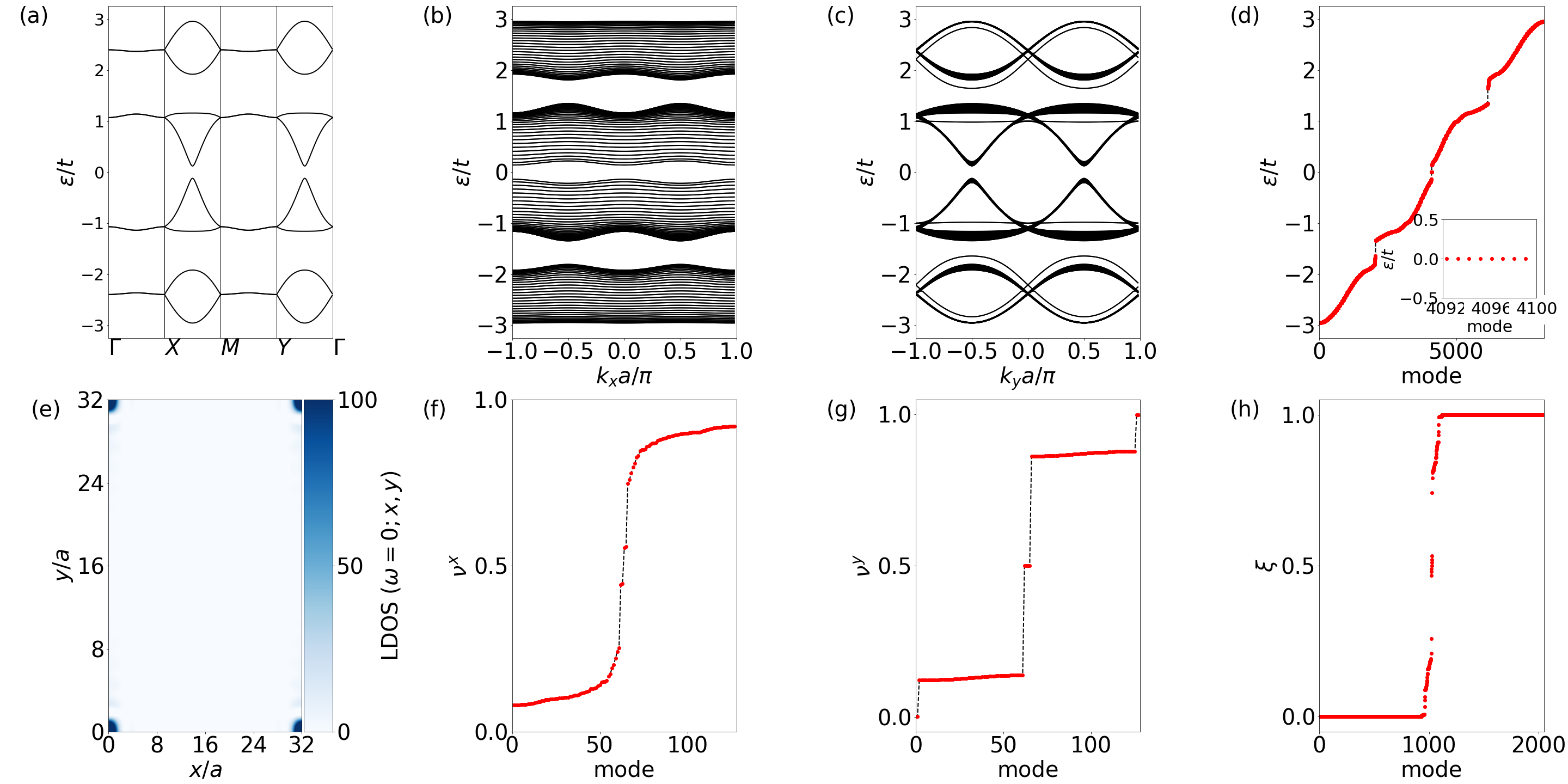}
    \caption{Main features of the hybrid phase. 
    Energy spectrum $\epsilon$ for fully periodic (a), $x$-periodic (b), $y$-periodic (c), and open (d) boundary conditions. Inset in (d) is a zoom-in on the modes in the middle of the spectrum with energy $\epsilon=0$. LDOS at zero energy (e). Wannier spectrum along $x$ (f) and $y$ (g). Entanglement spectrum $\xi$ (h). Parameters used: $B_x=0.9 t$, $\Delta=0.8 t$, $\lambda=0$, system size $32$ unit cells for directions with open boundary conditions and $100$ $k-$points.}
    \label{fig_HOTSC_II}
\end{figure*}

For small $B_x$ and $\Delta$, we find a phase that we call the HOTSC phase, indicated by the yellow region in Fig.~\ref{fig_phase}(b). The main features of this phase are summarized in Fig.~\ref{fig_HOTSC_I}. First, considering the energy $\epsilon$ spectrum for fully periodic in Fig.~\ref{fig_HOTSC_I}(a), $x$-periodic in Fig.~\ref{fig_HOTSC_I}(b), $y$-periodic in Fig.~\ref{fig_HOTSC_I}(c), and fully open in Fig.~\ref{fig_HOTSC_I}(d) boundary conditions, we realize that the bulk is gapped, while there are eight states at zero energy present only for fully open boundary conditions, Fig.~\ref{fig_HOTSC_I}(d).
The system is thus gapped under both $x$- and $y$-periodic boundary conditions in this phase. 
Plotting the LDOS at zero energy in Fig.~\ref{fig_HOTSC_I}(e), we see that these states are corner states, explaining why they appear just for open boundary conditions. To further understand the properties of these zero energy modes, we examine the Wannier spectrum $\nu$ along both $x$ in Fig.~\ref{fig_HOTSC_I}(f) and along $y$ in Fig.~\ref{fig_HOTSC_I}(g). The half-integer values of both $\nu_{x,y}$ indicate the higher-order character of this phase\footnote{We note here that with increasing the value of $\Delta$ and $B$, but still in the HOTSC phase, the midgap values of $\nu^x$ deviate from $0.5$ due to the breaking of mirror symmetry along $x$. Nevertheless, we still have robust corner modes with properties similar to the ones discussed below.}. The higher-order aspect is corroborated by the entanglement spectrum $\xi$ in Fig.~\ref{fig_HOTSC_I}(h), which shows distinct isolated half-quantized modes inside the midgap region.

Finding a total of eight corner modes, two at each corner, can be thought of as expected since the original BBH model at $B_x=\Delta=0$ host similar corner modes in this characteristic quadrupolar phase  \cite{PhysRevB.106.245109,Ghosh21}. To analyze these eight corner modes in our superconducting system, we consider the total (i.e.~the sum) Majorana polarization for the states at zero energy in Fig.~\ref{fig_HOTSC_I_modes}. We see that it is also localized in the corners of the system with a pattern that changes signs in the different corners, which is an important signature of MZMs \cite{ bena2017testing}. However, when computing $C$ (defined in Eq.~\eqref{eq_def_C} and using the grey shaded area in Fig.~\ref{fig_HOTSC_I_modes}), we obtain a much smaller than one (maximum of around $0.25$, with an overall increase with increasing $\Delta$). 
This indicates that, while these are zero-energy states, they should be classified as degenerate zero-energy Andreev bound states and not individual MZMs.
In summary, the HOTSC phase can be viewed as a superconducting extension of the quadrupolar phase of the BBH model at least regarding its surface states, but where the corner states are now a combination of particles and holes as they are Bogoliubov quasiparticles, but not still not MZMs.

\begin{figure}[!h]
    \includegraphics[width=0.75\linewidth]{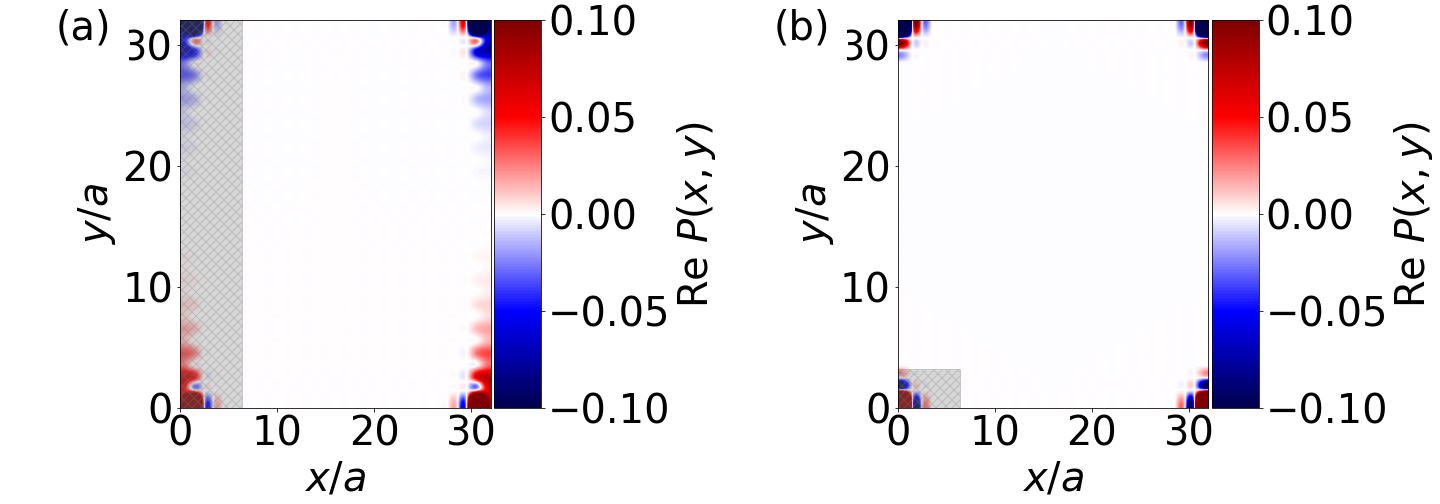}
    \caption{Sum of the Majorana polarizations $P$ of the hybrid phase boundary states. (a) represents this sum for the states that are localized in the edges, while (b) represents them sum for the corner states. The gray dashed area indicates the region where $C$ is calculated. For both the edge and corner states, $C=1$ (see text for discussion), indicating that these states are MZMs. Parameters same as in Fig.~\ref{fig_HOTSC_II}.
    }
    \label{fig_HOTSC_II_modes}
\end{figure}

\subsection{Hybrid phase}\label{subsec_HOTSC_II}
With increasing values of $\Delta$ and $B_x$, the system enters into what we name the hybrid phase, the green region in Fig.~\ref{fig_phase}(b). While present only in a narrow region of the phase diagram, it displays interesting features. 
\begin{figure*}[htb]
    \includegraphics[width=\linewidth]{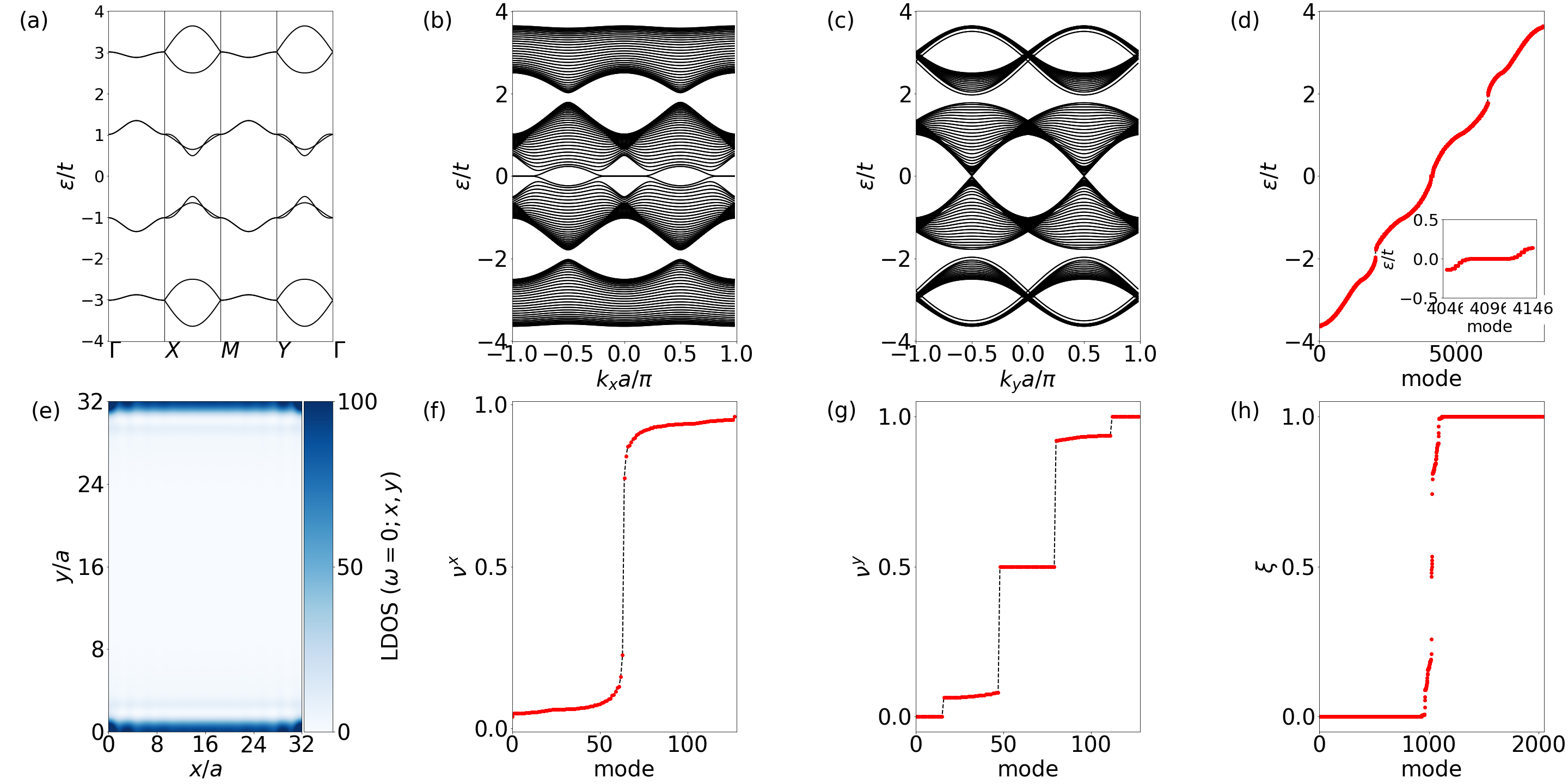}
    \caption{Main features of the nodal flat band phase. 
     Energy spectrum $\epsilon$ for fully periodic (a), $x$-periodic (b), $y$-periodic (c), and open (d) boundary conditions. Inset in (d) a zoom-in on the modes in the middle of the spectrum with energy $\epsilon=0$. LDOS at zero energy (e). Wannier spectrum along $x$ (f) and $y$ (g). Entanglement spectrum $\xi$ (h). Parameters used: $B_x=1.5 t$, $\Delta=0.9 t$, $\lambda=0$, system size $32$ unit cells for directions with open boundary conditions and $100$ $k-$points.}
    \label{fig_fb_maj_nodal}
\end{figure*}
We analyze the general properties of this phase in Fig.~\ref{fig_HOTSC_II}, with a similar set of data as for the HOTSC phase earlier. We find that it has energy spectrum similar to the HOTSC phase: both the bulk, Fig.~\ref{fig_HOTSC_II}(a), and the edges,  Figs.~\ref{fig_HOTSC_II}(b,c), are gapped while there are eight midgap states at zero energy for open boundary conditions,  Fig.~\ref{fig_HOTSC_II}(d). These midgap states are still located at the corners, as shown in the LDOS at zero energy, Fig.~\ref{fig_HOTSC_II}(e). However, when inspecting the Wannier spectrum, Figs.~\ref{fig_HOTSC_II}(f,g), we see that $\nu^y$ is still half-quantized, while $\nu^x$ exhibits a gap around $0.5$. Further, the entanglement spectrum in  Fig.~\ref{fig_HOTSC_II}(h) now shows a continuous array of midgap eigenvalues that are present symmetrically around  $\xi=0.5$. Therefore, this hybrid-order phase can be considered to be topologically distinct from the previous HOTSC phase. We remark that since $\nu^x$ is also not fully quantized for larger values of $\Delta$ and $B$ in the HOTSC phase, the topological phase transition between the HOTSC and the hybrid phases is due to the change of localization of some of the zero energy states, as we show below.

In fact, by going back to Figs.~\ref{fig_phase_PBC}(a,b), we see that there is no change in the energy gap from the phases HOTSC and hybrid for systems with periodic conditions along $x$, indicating that the change from the phase HOTSC to the hybrid phase is due to properties related to the I and III edges, along the $y$-direction. 
As a consequence, only when periodic boundary conditions are applied along the $y$-direction, the gap closing, indicating a topological phase transition, is noticed between the HOTSC and hybrid phases.

Although not very visible\footnote{The apparent localization just in the corners at zero energy is due to the stronger localization (hence, larger values of the wavefunction) of the corner states compared with states that spread over the edge.} in the total zero-energy LDOS in Fig.~\ref{fig_HOTSC_II} (e), we find that four, i.e.~half, of the zero-energy states in the hybrid phase now have a significant weight not just at the corners but also partially along the $y$-direction. To illustrate these localization properties better we plot the Majorana polarization divided up into two sets in Figs.~\ref {fig_HOTSC_II_modes}(a,b), respectively. Here it is now clear that four of the states are still corner states, just as in the HOTSC phase, but four other states are now substantially delocalized along the $y$-direction. We also find that 
both sign changing between different corners and $C=1$, provides strong indications that these states are now MZMs. We thus conclude that with all eight zero-energy modes now possible to spatially separate they can become MZMs. As a consequences, the increase of $\Delta$ and $B_x$ moving from the HOTSC to the hybrid phase turn the surviving four corner modes in the HOTSC phase corner-localized, single, MZMs and the other four turn into edge-extended MZMs.

\begin{figure}[htb]
    \includegraphics[width=0.75\linewidth]{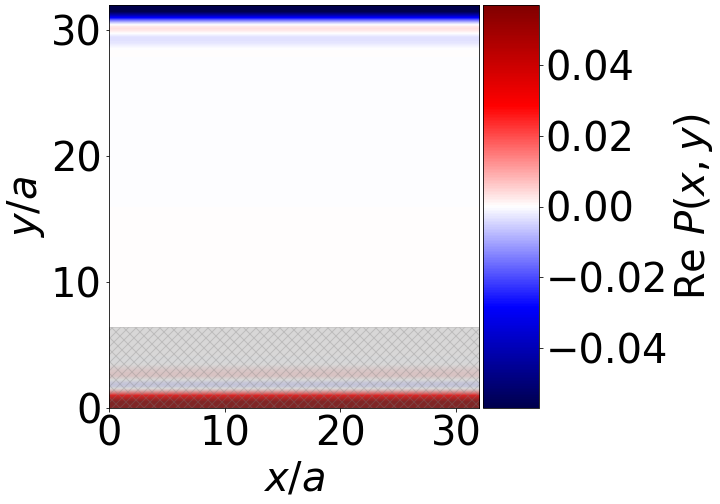}
    \caption{Sum of the Majorana polarizations $P$ for the four states with $\epsilon = 0$ for $k_x=0$ in the nodal flat band phase, using a partial Fourier transform, as explained in the main text. The gray dashed area indicates the region where $C$ is calculated. For all four states, $C=1$, indicating that they are MZMs. Parameters same as Fig.~\ref{fig_fb_maj_nodal}.}
    \label{fig_fb_maj_nodal_modes}
\end{figure}

In summary, we see that in this hybrid phase hosts a combination of edge and corner modes. Since there is no bulk gap closing (see Fig.~\ref{fig_phase_PBC} (a)) between the HOTSC and hybrid phase, we expect the change between the two phases to be caused by a change in the properties of the edge, which is similar to what happens in some other extrinsic higher-order phases \cite{PhysRevB.97.205135}. In fact, when considering the edge theory for this Hamiltonian in Sec.~\ref{subsec_corners}, we see that the magnetic field can change the mass profile at the edges, delocalizing some of the corner modes. The delocalized states are now instead localized at the edge and are still characterized by a quantized $\nu^y$. These states appear similar to the ones in the phases with polarization $p$ $p^x=0$ and $p^y=0.5$ (or vice-versa) in the original BBH model when the hopping along $x$ and $y$ is different \cite{benalcazar17_PRB, arouca2020thermodynamics}, named a dipolar phase in Ref.~\cite{arouca2020thermodynamics}, which is an example of a boundary obstructed atomic insulator \cite{PhysRevResearch.3.013239}. 
Due to the intriguing combination of corner and edge boundary modes, we name this a hybrid-order phase, as it has an inherent mix of different topologies.
However, we note that our use of the word hybrid should not be confused with the situation where more standard first- and second-order phases are appearing at the same time, also recently called a hybrid phase \cite{PhysRevLett.125.266804,Zhang2020,PhysRevLett.126.156801,Wang2023, hossain2024discovery}. Instead, our hybrid phase is a standard second-order phase appearing jointly with a dipolar phase. A clear difference is that in our hybrid phase the number of zero-energy boundary states remains constant (four corner modes and four edge modes), independent on system size, while any phase with a standard first-order character sees the number of edge modes grow with system size.
We remark that the only invariant we use that can distinguish this phase from the HOTSC phase is the nested entanglement spectrum, which shows a continuous array of values in contrast to the HOTSC phase where there are just modes at $0.5$.
\begin{figure*}[tb]
    \includegraphics[width=\linewidth]{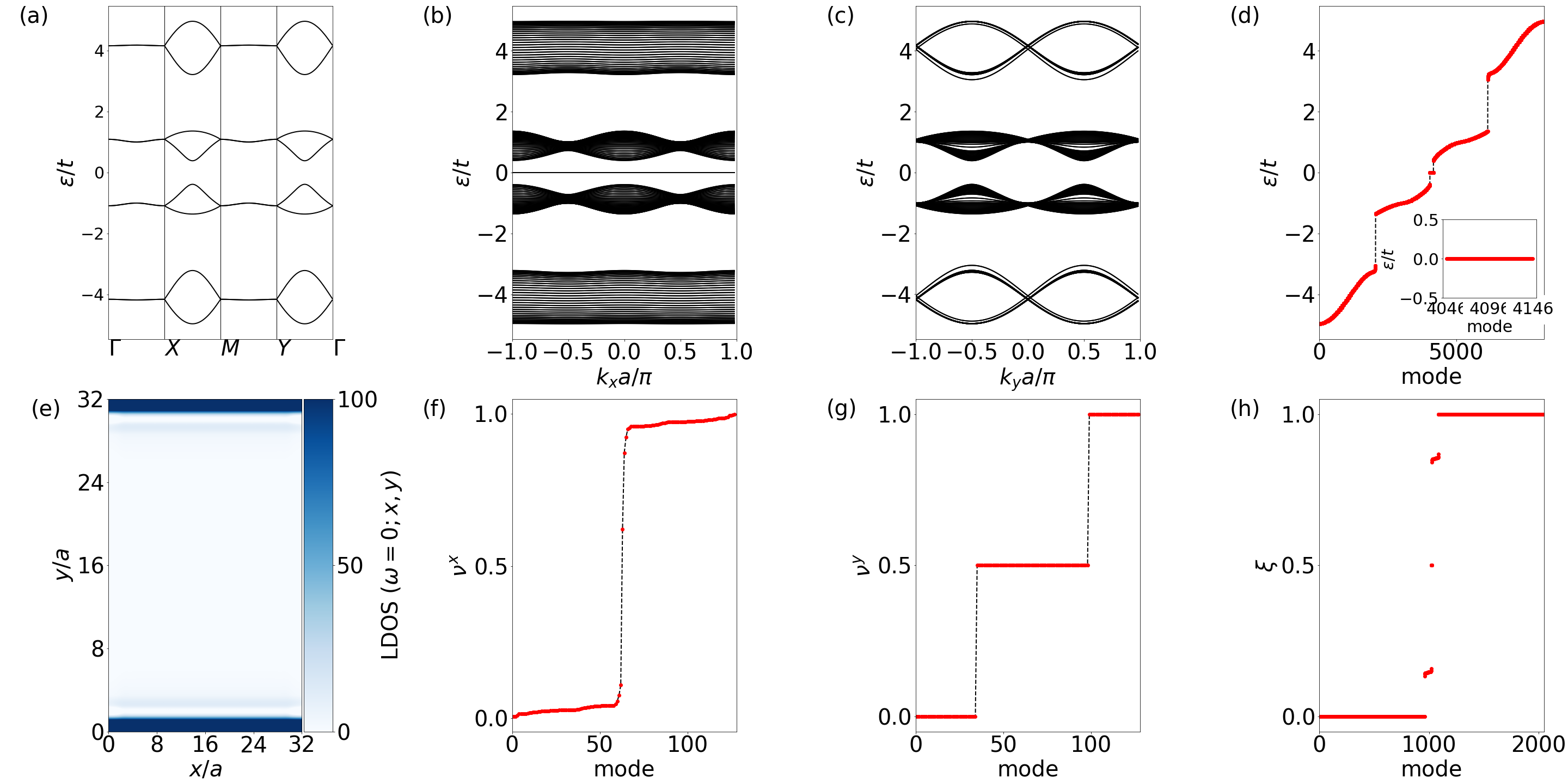}
    \caption{Main features of the phase with nodeless flat bands.
     Energy spectrum $\epsilon$ for fully periodic (a), $x$-periodic (b), $y$-periodic (c), and open (d) boundary conditions. Inset in (d) a zoom-in on the modes in the middle of the spectrum with energy $\epsilon=0$. LDOS at zero energy (e). Wannier spectrum along $x$ (f) and $y$ (g). Entanglement spectrum $\xi$ (h). Parameters used: $B_x=1.8 t$, $\Delta=2 t$, $\lambda=0$, system size $32$ unit cells for directions with open boundary conditions and $100$ $k-$points.}
    \label{fig_fb_maj_nodeless}
\end{figure*}

\subsection{Nodal flat bands phase}\label{subsec_nodal_FBM}
Increasing either $\Delta$ or $B_x$, we arrive at yet other phase, a nodal flat band phase, represented in blue in Fig.~\ref{fig_phase}(b). This phase is characterized by bulk nodal points, \textit{i.e.} there exists momenta where the bulk energy gap (superconducting gap) closes. Between these bulk nodal points, we find flat bands that appear localized to the boundaries of the system. We investigate the general properties of this phase in Fig.~\ref{fig_fb_maj_nodal}. Considering the energy dispersion along the high-symmetry points in Fig.~\ref{fig_fb_maj_nodal}(a), the bulk of the system seems to be gapped, in contradiction to Fig.~\ref{fig_phase_PBC}(a). However, analyzing the energy spectrum for the system with open boundary conditions along $y$, in Fig.~\ref{fig_fb_maj_nodal}(b) and $x$ in Fig.~\ref{fig_fb_maj_nodal}(c), we understand that this happens because the bulk gap is only zero at specific points $k_x a\neq 0, \pi$. The bulk gap thus vanishes away from the high-symmetry line and that is why the gap closing is not visible in Fig.~\ref{fig_fb_maj_nodal}(a). Additionally, there exist zero-energy flat bands along $k_x$, which connect these bulk nodal points, while all bands are dispersive along $k_y$. This suggests that the flat bands are boundary states localized along the edges II and IV in Fig.~\ref{fig_phase}(a). 
These flat bands cause the energy spectrum with fully open boundary conditions in Fig.~\ref{fig_fb_maj_nodal}(d) to host a macroscopic degeneracy at zero energy, as shown in the inset of Fig.~\ref{fig_fb_maj_nodal}(d).

Further evidence that the zero-energy states are localized along the II and IV edges is found in the LDOS at zero energy in Fig.~\ref{fig_fb_maj_nodal}(e), which shows weight just at these edges. This indicates that these are the boundary states of a first-order topological phase, since the absence of corner modes clearly discards these phase as any higher-order phase. We find this fully consistent with the topological invariants: an absence of midgap modes in $\nu^x$ in Fig.~\ref{fig_fb_maj_nodal}(f), while a number of 0.5 modes in $\nu^y$ appears in Fig.~\ref{fig_fb_maj_nodal}(g). We further find that the number of zero-energy states and 0.5 modes in $\nu^y$ scales with the number of unit cells along $x$, which further corroborates that this is an edge phenomenon along the $x$-direction. 
Here $\nu^y$ is thus an invariant that characterizes the presence of these zero-energy states, which is related by a bulk-boundary correspondence to the nodes in the bulk \cite{schnyder2008classification, matsuura2013protected, schnyder2015topological}. Another indication that this is not a second-order phase is the entanglement spectrum in Fig.~\ref{fig_fb_maj_nodal}, which does not show sharply quantized modes at 0.5 as for e.g.~the HOTSC phase. Instead, we find a discontinuous array of mid-gap eigenvalues symmetrically placed around 0.5.

To better understand the overall properties of the zero-energy flat bands, it is here most convenient to consider a system with open boundary conditions along $y$ and periodic boundary conditions along $x$ (corresponding to the spectrum of Fig.~\ref{fig_fb_maj_nodal}(b)) for $k_x=0$, finding four states with zero energy. After diagonalization, we then have a wavefunction $\tilde{\psi}(k_x=0, y)$. To obtain a complete real space wavefunction, we multiply $$\psi_{k_x=0}(x, y)= 1/\sqrt{L} \exp( i k_x x)\tilde{\psi} (k_x=0, y),$$  which is equivalent to a partial Fourier transform. This is the wavefunction we use to compute the Majorana polarization $P$ in Figs.~\ref{fig_fb_maj_nodal_modes}. We find that  $P$ changes sign between the two edges for all modes, and importantly we find $C=1$ when summing over the gray region, indicating that these edge states have Majorana properties. However, since there is a degeneracy of two zero-energy modes per edge per momentum in this phase, we cannot for sure classify these states as MZMs without considering if a linear combination of them cannot still recombine into complex fermionic modes. Still, because these modes both present a strong Majorana polarization and we have only two possible particle-hole symmetric pairs that we can build with both spin and orbital degrees of freedom, we choose call these states flat band MZMs.
We note that these flat band MZMs are extended along the edges, and their number increases with system size, in contrast to the MZMs obtained for the hybrid phase that are always just two per edge and one per corner, as discussed in Sec.~\ref{subsec_HOTSC_II}.

\subsection{Nodeless flat bands phase}\label{subsec_nodeless_FBM}
Finally, moving closer to the diagonal of the phase diagram, with $\Delta/t\approx B_x/t>1$, we find a nodeless flat band phase, red in Fig.~\ref{fig_phase}(b). For this phase, the bulk system is gapped, as seen in Fig.~\ref{fig_fb_maj_nodeless}(a), as is the semi-infinite spectrum with the open boundary along $x$, illustrated in  Fig.~\ref{fig_fb_maj_nodeless}
(c). In Fig.~\ref{fig_fb_maj_nodeless}(b), we instead find zero-energy flat bands spanning the whole Brillouin zone along $k_x$. Fully open boundary conditions also generate zero-energy mid-gap states in Fig.~\ref{fig_fb_maj_nodeless}(d).  Plotting the zero-energy LDOS we find that these flat bands are localized on the II and IV edges in Fig.~\ref{fig_fb_maj_nodeless}(e). Further, although $\nu^x$ is completely gapped in Fig.~\ref{fig_fb_maj_nodeless}(f), $\nu^y$ in Fig.~\ref{fig_fb_maj_nodeless}(g) has a finite number of $0.5$ modes, which grows with the number of unit cells along $x$. This $\nu^y$ profile confirms that the zero-energy states are an edge phenomenon along the $x$-direction.  Finally, the entanglement spectrum in Fig.~\ref{fig_fb_maj_nodeless}(h) shows a discontinuous array of mid-gap eigenvalues around 0.5, gathering reasonably close to 0.5. This is qualitatively similar to the nodal flat band case, however, the discontinuous profile is more asymmetric around 0.5 in this case compared to the previous case. 

\begin{figure}[htb]
    \includegraphics[width=0.75\linewidth]{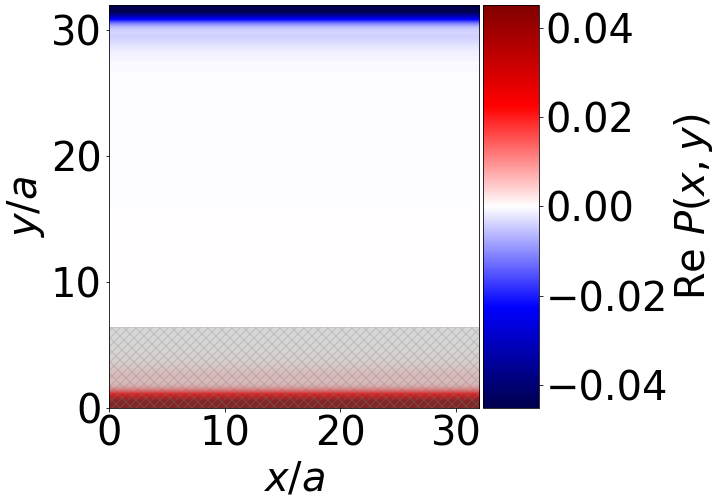}
    \caption{Sum of the Majorana polarizations $P$ for the four states with $\epsilon = 0$ for $k_x=0$ in the nodeless flat band phase, using a partial Fourier transform, as explained in the main text. The gray dashed area indicates the region where $C$ is calculated. For all four states, $C=1$, indicating that they are MZMs. Parameters same as Fig.~\ref{fig_fb_maj_nodeless}.}
    \label{fig_fb_maj_nodeless_modes}
\end{figure}

We investigate the Majorana polarization of the states at zero energy in this phase in Fig.~\ref{fig_fb_maj_nodeless_modes}. We here again focus on the semi-periodic system with $k_x=0$ (corresponding to the spectrum of Fig.~\ref{fig_fb_maj_nodeless} (b)) as done for the nodal phase. Although the modes present a similar spatial profile, we remark that these flat bands are here present without any corresponding nodal points in the bulk. Instead we have a flat zero-energy edge state entirely separated from the bulk gap spectrum.
 As such, they are a very different kind of surface state compared to those of the nodal flat band phase. 
We currently actually do not know how they can be fully classified in terms of symmetry-protected topological phases. Nevertheless, we still know that they are topologically protected by symmetry since the number of states at zero energy is determined by the number of $0.5$ states in $\nu^y$, which is a topological invariant. 
Finally, we note that these flat bands states present $C=1$, which makes us designate them as flat band MZMs, in a similar way to in the nodal flat band case.

%%%%%%%%%%%%%%%%%%%%%%%%%%%
% SECTION: ANALYTICAL
%%%%%%%%%%%%%%%%%%%%%%%%%%%
\section{Analytical results}\label{sec_disc}
After having described in detail the properties of all the different topological phases numerically, we also present some results feasible to achieve analytically in order to enhance the overall understanding. While a complete analytical treatment at present seems not feasible, we can still obtain selective analytical results. Here, we first present analytical calculations supporting the existence of a nodal phase despite an $s$-wave superconducting order parameter. For this purpose, we define a dual Hamiltonian. In addition to a dual Hamiltonian, we also extract a low-energy Hamiltonian and are able to show analytically that corner states exist in its regime of validity. 

\subsection{Nodal superconductivity}\label{subsec_gap}
One of the phases, the nodal flat band phase, depicted in blue in Fig.~\ref{fig_phase}(b) and discussed in detail in Sec.~\ref{subsec_nodal_FBM}, hosts bulk nodes. This is despite the model only containing on-site, or in $k$-space isotropic $s$-wave, pairing, which are usually associated with a fully gap, while nodes are usually thought of as requiring a $k$-dependent order parameter. The presence of nodes in the nodal flat band phase can be understood as a consequence of the spin-orbit coupling in the normal state Hamiltonian, as such terms can induce an effective momentum-dependent pairing. The equivalence between the spin-orbit coupling and nodal superconductivity can be seen explicitly by performing the unitary transformation of the Hamiltonian of Eq.~\eqref{eq_h_BdG} (for $\lambda=0$):
\begin{equation}
    h_D(\mathbf{k})=U h_{\rm BdG}(\mathbf{k}) U^\dagger; \, \, U=\begin{pmatrix}\sigma_0\otimes s_0& i\sigma_2\otimes s_2\\ i \sigma_2\otimes s_2&\sigma_0\otimes s_0\end{pmatrix},
\end{equation}
where we obtain a dual Hamiltonian $h_D$, following the nomenclature of Ref.~\cite{sato09dual}, with the form
% \begin{widetext}
\begin{eqnarray}
    \hspace{-0.5cm} h_D(\mathbf{k})&=&\tau_1\big[-t\sin(k_ya)\sigma_0 s_3-t \cos(k_x a)\sigma_0 s_1 \nonumber \\
    \hspace{-0.5cm} &+& B_x \sigma_2s_3 \big]
    +\Delta \tau_3\sigma_1s_0-t\cos(k_y) \tau_0\sigma_2s_2,
    \label{eq_dual}
\end{eqnarray}
% \end{widetext}
where we omit the outer product symbol $\otimes$  for convenience. 

In the dual Hamiltonian in Eq.~\eqref{eq_dual}, the superconducting pairing appears in multiple terms. There is pairing with $p$-wave intra-orbital spin-triplet  pairing symmetry given by $-t\sin(k_y a) \tau_1\otimes \sigma_0\otimes s_3$, with extended $s$-wave intra-orbital spin-singlet symmetry given by $-t \cos(k_x a)\tau_1\otimes \sigma_0\otimes s_1$, and with $s$-wave odd-interorbital spin-triplet symmetry given by $B_x \tau_1\otimes \sigma_2\otimes s_3$. The first two terms always present nodes at some $k_x$ and $k_y$ coordinates and result in the nodal profile of the superconducting order parameter. If such nodes in the superconducting order parameter also overlap with the normal-state Fermi surface, then the system will have a nodal energy gap. In the presence of the third term $B_x \ne 0$, the nodal structure caused by the first two terms can still be retained with nodes then appearing at different values of $k_x$ and $k_x$ in appropriate parameter regimes. This qualitatively explains why a nodal state at all can be possible and also shows how it is intricately linked to the $\Delta$ and $B_x$ parameters. We further note that in the pairing terms, the dependence on $k_x$ comes from a hopping term $\cos(k_x a)$, while $\sin(k_y a)$ appears due to the spin-orbit coupling of the normal state. This difference, together with the fact that the magnetic field is applied along $x$, explains the asymmetry seen in the properties of the system with periodic boundary conditions along $x$ or $y$. Finally, the existence of a $p$-wave term is also an underlying reason for the existence of MZMs in several of the different phases. 

\subsection{Corner states}\label{subsec_corners}
In addition to the dual Hamiltonian, we can also analytically study a low-energy continuum  Hamiltonian. This is a technique commonly used to connect the presence of localized boundary states with a non-trivial mass profile, as first constructed by Jackiw and Rebbi\cite{PhysRevD.13.3398}, and where the mass profile tells us some properties of the topological phase. For instance, for a higher-order topological system, the mass profile needs to change sign between two adjacent edges \cite{benalcazar17_PRB, Ghosh21}, with the corner in between hosting the localized boundary state. To obtain the most simple low-energy Hamiltonian, we keep just terms that are first-order in momentum in $h_{\rm BdG}$ in Eq.~\eqref{eq_h_BdG} and obtain 
% \begin{widetext}
\begin{equation}
    h_\Gamma(\mathbf{k})=t k_y a \Gamma_1+t\Gamma_2+t k_x a \Gamma_3+t \Gamma_4+\Delta \Gamma_5+B_x\Gamma_6.
    \label{eq_h_Gamma}
\end{equation}    
% \end{widetext}
Here $\mathbf{k}=(k_x, k_y)$ represents the continuum momentum in relation to the $\Gamma$ point. 

We can obtain the low-energy descriptions of any corner states of $h_\Gamma$ by substituting $k_{x/y}\rightarrow -i \partial_{x/y}$ and looking for localized solutions. The complete calculation is reported in detail in Appendix~\ref{app_edge}, and here we comment on the results. Considering appropriate boundary conditions, the solutions of $h_\Gamma$ localized on the I and III edges are 
\begin{equation}
    \psi^{\textrm{I}/\textrm{III}}(x, y)=\sum\limits_{l, m=\pm 1}c_{l,m}\mathcal{N}_x e^{\mp \sqrt{2} \frac{x}{a}} e^{i k_y y} \chi_{l, m}^{\textrm{I}/\textrm{III}},
\end{equation}
where $\mathcal{N}_x$ is a normalization factor, $\sum \left|c_{l,m}\right|^2=1$, and the spinors $\chi_{l,m}^{\textrm{I}/\textrm{III}}$ are given by
\begin{equation}
    \chi_{l,m}^{\textrm{I}}=\ket{l, m} \frac{\ket{s=1}+\left(m-\sqrt{2}\right)\ket{s=-1}}{\sqrt{4-2m\sqrt{2}}},
    \label{eq_chi_I}
\end{equation}
and 
\begin{equation}
    \chi_{l,m}^{\textrm{III}}=\ket{l,m} \frac{\ket{s=1}-\left(m+\sqrt{2}\right)\ket{s=-1}}{\sqrt{4+2m\sqrt{2}}},
    \label{eq_chi_III}
\end{equation}
where $l=\pm 1$, $m=\pm 1$, and $s=\pm 1$ are eigenvalues of $\tau_3$, $\sigma_3$ and $s_3$, respectively. 
We further obtain the edge Hamiltonians by projecting the low-energy Hamiltonian in the subspace of Eqs.~\eqref{eq_chi_I} and \eqref{eq_chi_III}, yielding
\begin{eqnarray}
    h^{\textrm{I}}(k_y)&=&-t k_y a\, \tau_3 \otimes \sigma_2-\frac{B_x}{\sqrt{2}}\,\tau_3\otimes \sigma_0\label{eq_ham_surf_I},\\
    h^{\textrm{III}}(k_y)&=&-t k_y a\, \tau_3 \otimes \sigma_2-\frac{B_x}{\sqrt{2}}\,\tau_3\otimes \sigma_0\label{eq_ham_surf_III}.
\end{eqnarray}
These are Dirac-like Hamiltonians with a mass term proportional to $B_x$. Note that both solutions in Eqs.~\eqref{eq_chi_I} and \eqref{eq_chi_III} are polarized in the effective particle-hole space, represented by $\tau_3$,  while they are a combination of different spin states with the effective spin degrees of freedom represented by $\sigma_3$. Note here that since the edge Hamiltonians are projected Hamiltonians, the original notion of particle-hole or spin is replaced by their effective notions.    
Turning to the localized solutions on the II and IV edges, we obtain similarly as on the other edges
\begin{equation}
    \psi^{\textrm{II}/\textrm{IV}}(x, y)=\sum\limits_{l,m=\pm 1}\tilde c_{l,m}\mathcal{N}_y e^{i k_x x}e^{\mp \sqrt{2} \frac{y}{a}} \chi_{l,m}^{\textrm{II}/\textrm{IV}},
\end{equation}
where $\mathcal{N}_y$ is a normalization factor, we have $\sum \left|\tilde c_{l,m}\right|^2=1$, and the spinors $\chi_{lm}^{\textrm{II}/\textrm{IV}}$ are given by
\begin{equation}
    \chi_{l,m}^{\textrm{II}}=\ket{l, m} \frac{\ket{s=1}-m\left(1+\sqrt{2}\right)\ket{s=-1}}{\sqrt{4+2\sqrt{2}}},
    \label{eq_chi_II}
\end{equation}
and 
\begin{equation}
    \chi_{l,m}^{\textrm{IV}}=\ket{l, m} \frac{\ket{s=1}+m\left(1-\sqrt{2}\right)\ket{s=-1}}{\sqrt{4+2\sqrt{2}}},
    \label{eq_chi_IV}
\end{equation}    
where again $l=\pm 1$, $m=\pm 1$, and $s=\pm 1$ are eigenvalues of $\tau_3$, $\sigma_3$ and $s_3$, respectively. 

Again projecting on the solutions of Eqs.~\eqref{eq_chi_II} and \eqref{eq_chi_IV}, we obtain the effective Dirac Hamiltonians
\begin{eqnarray}
    h^{\textrm{II}}(k_x)&=&t k_x a \,\tau_3 \otimes \sigma_2-\frac{1}{\sqrt{2}}B_x\,\tau_3\otimes \sigma_3,\label{eq_ham_surf_II}\\
    h^{\textrm{IV}}(k_x)&=&t k_x a\, \tau_3 \otimes \sigma_2-\frac{1}{\sqrt{2}}B_x\,\tau_3\otimes \sigma_3\label{eq_ham_surf_IV}.    
\end{eqnarray}

Having extracted all four edge Hamiltonians, Eqs.~\eqref{eq_ham_surf_I}, \eqref{eq_ham_surf_III}, \eqref{eq_ham_surf_II}, and \eqref{eq_ham_surf_IV}, we note that all four edges have a mass term proportional to $B_x$. On the I and III edges, it is proportional to $\tau_3\otimes \sigma_0$, while on the II and IV edges it is proportional to $\tau_3\otimes \sigma_3$. Since eigenvalues of $\sigma_0$ and $\sigma_3$ can be different in terms of their signs, there exists a sign change in the mass term for adjacent edges. For a given subspace defined by $\tau_3$, we can hence obtain a positive mass $B_x$ on edges II and IV, while having a negative mass $-B_x$ on edges I and III, when $\sigma_3$ is projected on the effective spin-down states. This leads to the formation of corner modes between adjacent edges. On the other hand, for the effective spin-up projections, all the edges have negative mass terms given by $-B_x$, and hence corner modes are not expected to appear. The effective spin-polarized nature of the corner modes can be attributed to the sign-polarized Majorana polarization for a given corner mode. This analysis qualitatively explains the emergence of corner states for our model, Eq.~\eqref{eq_h_BdG}, in both the HOTSC and hybrid phases.

%%%%%%%%%%%%%%%%%%%%%%%%%%%
% SECTION: CONCLUSIONS
%%%%%%%%%%%%%%%%%%%%%%%%%%%
\section{Conclunding remarks}\label{sec_conc}

In this work, we investigate the topological phases of an orbital BBH model proximitized to a conventional spin-singlet $s$-wave superconductor and in the presence of an in-plane magnetic field. The interpretation of the BBH model in terms of orbital and spin degrees of freedom makes the superconducting pairing have a matrix structure not present in the original dimerized lattice of the BBH model and allows for an intriguingly rich topological phase diagram, despite using only a conventional superconductor. 
We map out the resulting phase diagram by considering different boundary conditions and investigate the topology of each phase by calculating both the Wannier and entanglement spectra, as well as the Majorana polarization.

At weak superconducting pairing $\Delta$ and magnetic field $B_x$, we find a HOTSC phase (yellow in Fig.~\ref{fig_phase}(b)) with eight zero-energy corner modes. This phase can be seen as analogous to the standard second-order topological phase in the BBH model, at least when considering the surface states, although the latter now also have particle and hole character since the system is superconducting, but still not being MZMs. Beyond this expected HOTSC phase, we also find several other, much more unexpected, topological phases. First, an unusual hybrid phase (green in Fig.~\ref{fig_phase}(b)) presents an atypical mix between a second-order and a dipolar topological phase. Here four zero-energy corner states from the HOTSC phase are preserved as expected for a second-order topological phase, while a dipolar phase contributes another four zero-energy edge states, all experiencing MZM character. Notably, the number of edge states does not grow with system size in this hybrid phase, but the number of zero-energy states remains fixed, at eight states. 
Second, two additional first-order phases present symmetry-protected zero-energy flat bands on opposite edges, with either nodal or nodeless bulk dispersion. The nodal flat bands phase (blue in Fig.~\ref{fig_phase}(b)) presents flat bands MZMs that are straightforwardly connected by a bulk-boundary correspondence to the bulk nodes. However, in the nodeless flat bands phase (red in Fig.~\ref{fig_phase}(b)) the bulk surprisingly remains fully gapped and the flat bands MZMs now spans the whole edge Brillouin zone and are protected by a quantized Wannier spectrum.
These results not only establish the rich phase diagram of a superconducting BBH model, but also, importantly demonstrate the existence of an unusual hybrid mixing of topological phases and shows that a zero-energy edge flat band can exists also for a nodeless, i.e.~fully gapped, bulk, such that the edge states are not continuously connected to the bulk bands.

In terms of experimental feasibility of the BBH superconductor model developed in this work, we can refer to the atom-optics setup for the realization of lattice tight-binding topological models  \cite{meier2016observation}. Apart from metamaterials \cite{Noh2018, Experiment3DHOTI.aSonicCrystals, Schulz2022}, higher-order topological insulator phases have also been experimentally observed in van der Waals stacking of bismuth-halide \cite{noguchi2021evidence}, as well as 
Bi$_x$Sb$_{1-x}$ alloys \cite{aggarwal2021evidence}.
The $s$-wave superconductor can be, in principle, placed in proximity to the above materials with appropriate substrates, such that HOTSC phases are realized in the above material. It is worth noting that superconductivity can be induced   
in the surface states of Bi$_{2}$Se$_{3}$, HgTe etc.~via the proximity effect~\cite{Experiment3DTIProximity1,PhysRevB.96.165302,Experiment3DTIProximity2}, with a superconducting gap even around $\Delta_{0}\sim ~\rm 0.5~meV$~\cite{Experiment3DTIProximity2}, despite the fully gapped bulk. Given the above experimental developments in the solid-state topology, we believe that the HOTSC phases obtained in our work are at least within future experimental reach.

We further note that we study the HOTSC model starting from a higher-order topological insulator limit of the underlying BBH model.  It would also be interesting to study a BBH superconductor model starting from a metallic bulk, where by tuning the chemical potential different topological phases may be uncovered in addition to those found in this work. One can hence investigate the three-parameter interplay between chemical potential, superconducting gap function, and magnetic field for engineering of HOTSC phases.  Moreover, the possibility to alternate between topological states with different spatial localizations by tuning the magnetic field may lead to interesting applications where these states are useful, including topological quantum computation.

{\it Note added:} During the final preparation of this work we became aware of the recent work Ref.~\cite{lapp2023complete}, where zero-energy flat bands are also found to appear as topological boundary states in a nodeless superconductor. However, in that case the flat band is protected by a bulk invariant in three dimensions (3D). Our flat band instead appears in a 2D system and is protected by a 1D edge invariant. It remains to be studied if deeper similarities between the two exist.

\begin{acknowledgments}
We are grateful for discussions with  A.~K.~Ghosh, A.~Saha, S.~Mondal, B.~Roy, V.~Juricic, O.~A.~Awoga, D.~Chakraborty, and A.~Bhattacharya. We acknowledge financial support from the Knut and Alice Wallenberg Foundation. Part of the simulations were enabled by resources provided by the National Academic Infrastructure for Supercomputing in Sweden (NAISS) and the Swedish National Infrastructure for Computing (SNIC) at UPPMAX, partially funded by the Swedish Research Council through grant agreements no.~2022-06725 and no.~2018-05973.
\end{acknowledgments}

%%%%%%%%%%%%%%%%%%%%%%%%%%%
% SECTION: APPENDIX
%%%%%%%%%%%%%%%%%%%%%%%%%%%
\appendix

\section{Finite $\lambda$}\label{app_phase_finite_lambda}

\begin{figure}[!htb]
    \includegraphics[width=\linewidth]{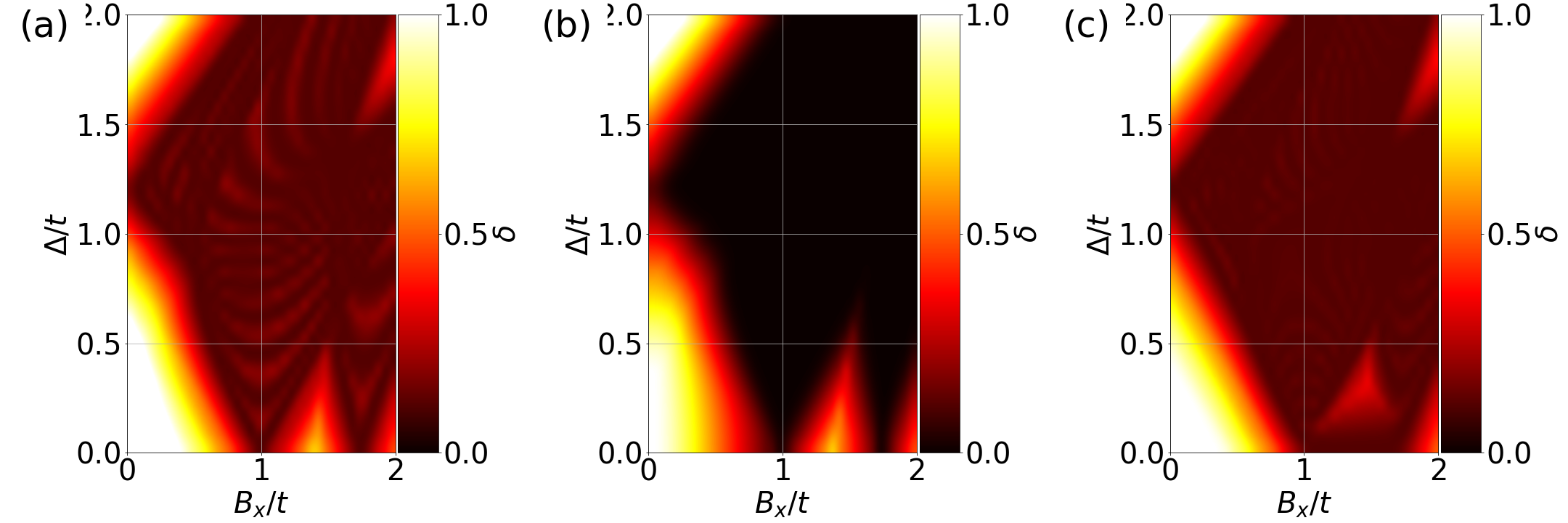}
    \caption{Energy gap $\delta$ derived from Eq.~\eqref{eq_h_BdG} as a function of $\Delta/t$ and $B_x/t$ for (a) fully-periodic, (b) $x$-periodic (open in $y$) and (c) $y-$periodic (open in $x$) boundary conditions. Parameters used: $\lambda=0.5t$, system size $32$ unit cells in each direction.}
    \label{fig_phase_05}
\end{figure}

In the main text, we focus on the properties of Eq.~\eqref{eq_h_BdG} for $\lambda=0$. Here, we discuss the effect of having a finite $\lambda$ in the phases discussed in the main text. For large $\lambda>t$, the system is just a trivial superconductor. For smaller but finite $\lambda$, the phase boundaries of the system change considerably but the topological phases largely persist. In Fig.~\ref{fig_phase_05} we set $\lambda = 0.5t$ and we repeat the plot of the energy gap $\delta$ as a function of $B_x$ and $\Delta$ from Fig.~\ref{fig_phase_PBC}. We still retain the phases discussed in the main text, although the phase boundaries change considerably and now cannot be computed analytically. In addition, we obtain a new phase, which we call the dipolar phase, described next.

\begin{figure*}[!htb]
    \includegraphics[width=\linewidth]{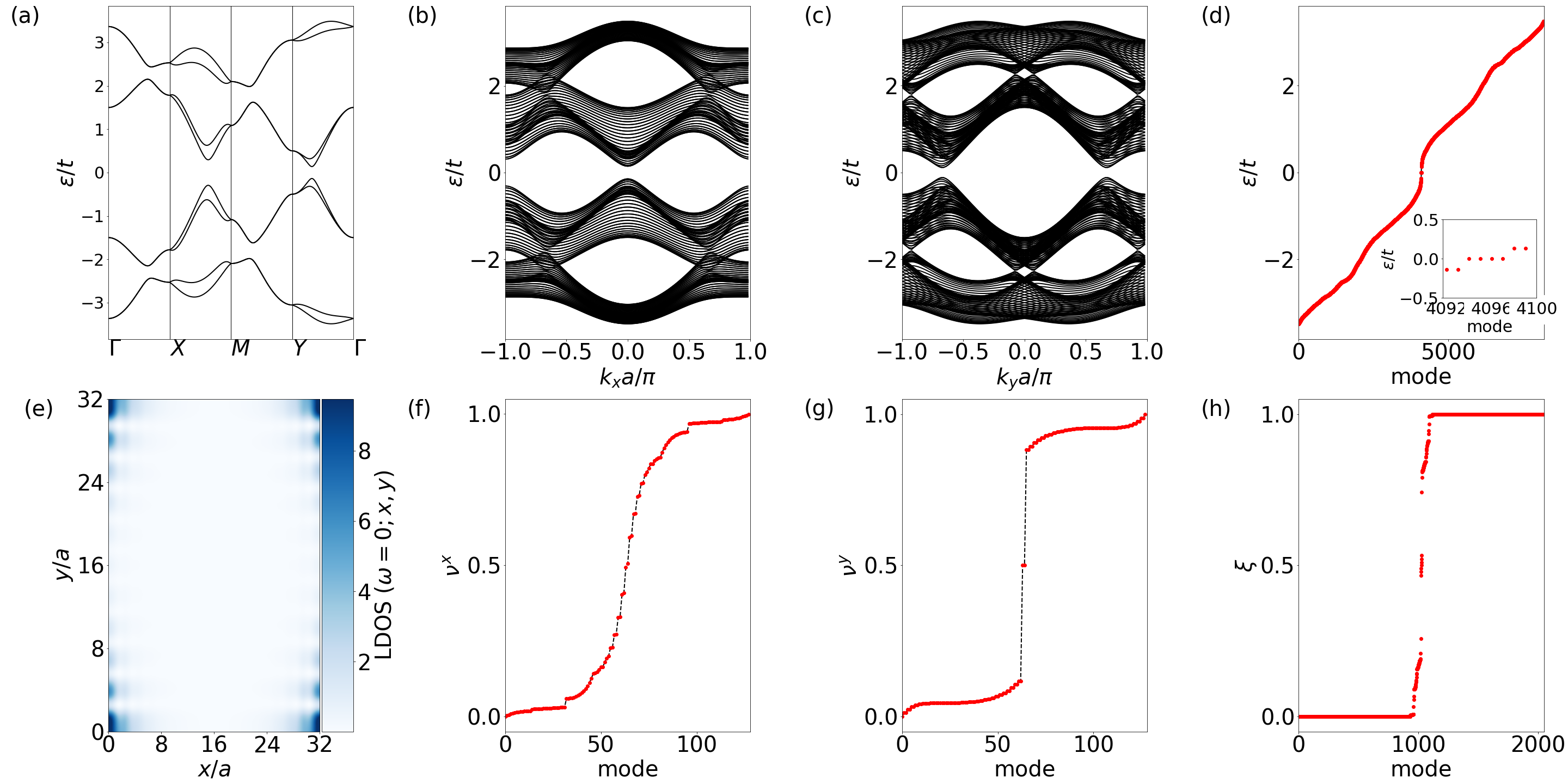}
    \caption{Main features of the dipolar phase.  Energy spectrum $\epsilon$ for fully periodic (a), $x$-periodic (b), $y$-periodic (c), and open (d) boundary conditions. Inset in (d) a zoom-in on the modes in the middle of the spectrum with energy $\epsilon=0$. LDOS at zero energy (e). Wannier spectrum along $x$ (f) and $y$ (g). Entanglement spectrum $\xi$ (h).. Parameters used: $B_x=1.5t$, $\Delta=0.2t$, $\lambda=0.5$, system size $32$ unit cells for directions with open boundary conditions and $100$ $k-$points.
    \label{fig_dip}}
\end{figure*}
\begin{figure}[!htb]
    \includegraphics[width=0.75\linewidth]{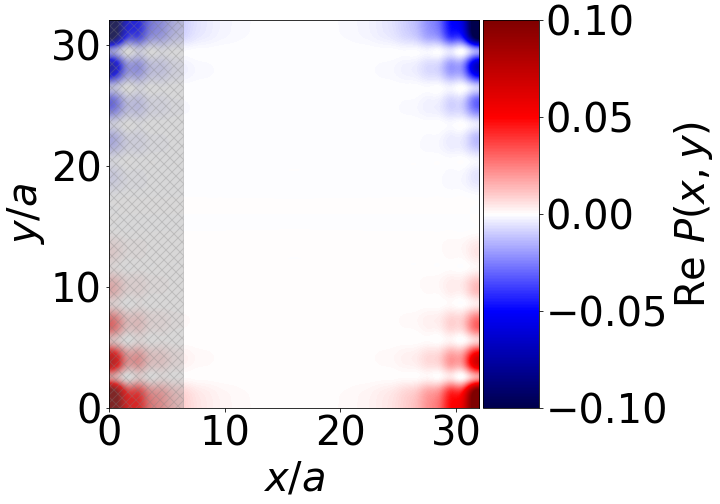}
    \caption{Dipolar phase boundary states features. Same panels as Fig.~\ref{fig_HOTSC_I_modes} for the four states with smallest (negative energy), but where only the last two states (c,d,g,h) have energy $\epsilon=0^-$, in the dipolar phase. Parameters same as in Fig.~\ref{fig_dip}.}
    \label{fig_dip_modes}
\end{figure}

\subsection{Dipolar phase}\label{app_dip}
With $\lambda$ finite, but smaller than $t$, we find a different kind of first-order topological phase appearing for higher values of either $\Delta$ and $B_x$, when one is much larger than the other. This phase, as shown in Fig.~\ref{fig_dip}(a-d), has a gapped bulk but with some localized midgap states on the I and III edges, as also shown in the zero-energy LDOS in Fig.~\ref{fig_dip}(e). The $\nu^x$ invariant along $x$ is not quantized, see Fig.~\ref{fig_dip}(f), while the $\nu^y$ presents two half-quantized values, see Fig.~\ref{fig_dip}(g), indicating that there are four symmetry-protected modes. The entanglement spectrum shows a symmetric distribution of eigenvalues excluding 0.5 within the mid gap region, as depicted in Fig.~\ref{fig_dip}(h).

In Fig.~\ref{fig_dip_modes}, we show the sum of the Majorana polarization for these states at zero energy. We see that they are localized on the edges of the system and also present $C=1$, indicating that they are MZMs. These states are very similar to the edge localized, or dipolar, states that appear in the hybrid phase, see Fig.~\ref{fig_HOTSC_II_modes}(a). Moreover, this phase resembles (a superconducting version of) the phase with $p^x=0$ and $p^y=0.5$ in the BBH model \cite{benalcazar17_PRB} called dipolar phase in Ref.~\cite{arouca2020thermodynamics}, such that we also call it a dipolar phase. We note that the $P$ of the zero-energy boundary states appropriately also resembles an electric dipole.

\begin{figure*}[!htb]
    \includegraphics[width=\linewidth]{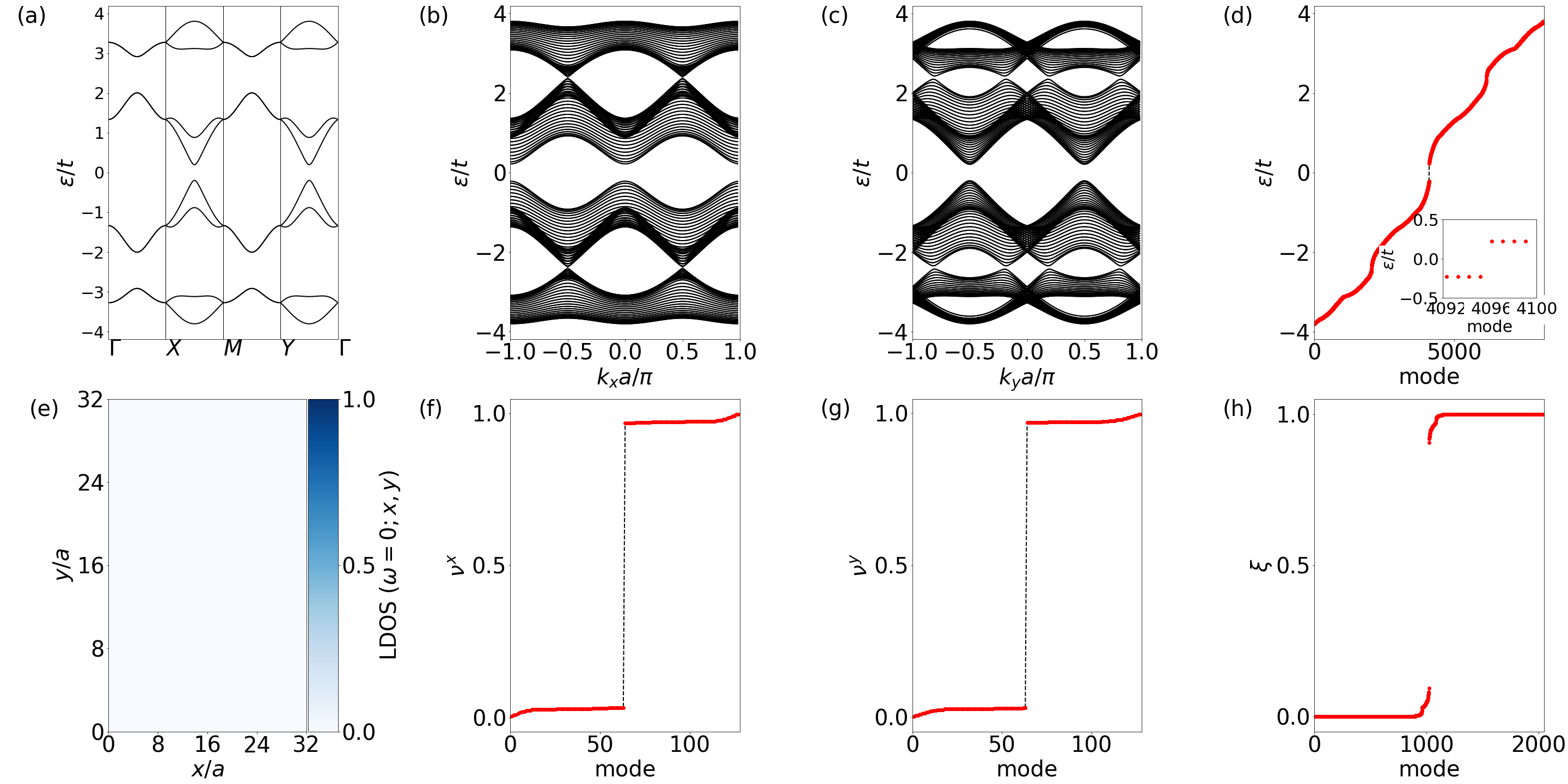}
    \caption{Main features of the trivial phase.   Energy spectrum $\epsilon$ for fully periodic (a), $x$-periodic (b), $y$-periodic (c), and open (d) boundary conditions. Inset in (d) a zoom-in on the modes in the middle of the spectrum with energy $\epsilon=0$. LDOS at zero energy (e). Wannier spectrum along $x$ (f) and $y$ (g). Entanglement spectrum $\xi$ (h). Parameters used: $B_x=2 t$, $\Delta=0.5 t$, $\lambda=0$, system size $32$ unit cells for directions with open boundary conditions and $100$ $k-$points.}
    \label{fig_trivial}
\end{figure*}

\section{Trivial phase}\label{app_triv}

For high values of only $\Delta$ or $B_x$, we find a topologically trivial phase. For $\lambda=0$, this occurs for $\Delta>\sqrt{2t^2+B_x\left(2t+B_x\right)}$ or $\Delta<-t+\sqrt{B_x^2-t^2}$, corresponding to the white region in Fig.~\ref{fig_phase}(b). Although this phase is topologically trivial, for completeness, we display the main features of this phase in Fig.~\ref{fig_trivial}. The system is gapped for all boundary conditions, see Figs.~\ref{fig_trivial}(a-d), which is supported by the LDOS at zero energy, see Fig~\ref{fig_trivial}(e), which shows no occupation. Further, all topological invariants indicate a trivial system, see Figs.~\ref{fig_trivial}(f-h).

\section{Derivation of edge Hamiltonians}\label{app_edge}
Here we provide the detailed derivation of the edge Hamiltonian expressions used in the main text. We continue in the deep topological limit of the BBH model with $\lambda=0$. Since the continuum Hamiltonian Eq.~\eqref{eq_h_Gamma} is composed of linear operators in $x$, $y$, and matrices acting on the particle-hole $\tau$, orbital $\sigma$, and spin $s$ degrees of freedom, we can write a generic wavefunction as 
\begin{equation}
    \psi(x,y)=\sum\limits_\alpha c_\alpha \phi^\alpha(x) \varphi^\alpha(y) \chi_\alpha,
\end{equation}
where $\phi^\alpha$ ($\varphi^\alpha$) is a complex spatial function of $x$ ($y$), and $\chi_\alpha$ are eight component spinors labelled by $\alpha$. Using this form, we can obtain solutions that are localized on specific edges (or corners) by checking how the Hamiltonian acts individually in each part.  

For the I and III edges, we look for localized solutions on $x$. We split $h_\Gamma=h_{0x}+h_{ky}$, where $h_{0x}=t\Gamma_2+t k_x a \Gamma_3+t \Gamma_4$ is the part which determine the zero states localized along $x$, while we treat $h_{ky}=tk_y a \Gamma_1+\Delta \Gamma_5+B_x\Gamma_6$ as its perturbation, reasonable given the fact that $B_x,\Delta \ll t$. Next we perform the substitution $k_x\rightarrow -i \partial_x$ and look for solutions of 
% \begin{widetext}
    \begin{equation}
    h_{0x}\psi(x, y)=\left[t \Gamma_2-i t \partial_x a \Gamma_3+t \Gamma_4\right]\psi(x,y)=0.
\end{equation}

Using the ansatz $\phi_\alpha(x)=\exp(q_x x/a)$ and that the $\chi_\alpha$ are linearly independent, we get the matrix equations
% \begin{widetext}
\begin{equation}
    \left[t \tau_3\sigma_2s_2-i t q_x \tau_3\sigma_2s_3+t \tau_3\sigma_1s_0\right]\chi_\alpha=0,
\end{equation}
% \end{widetext}
where we omit the external product for convenience.  Multiplying the whole equation by $\tau_3\otimes \sigma_2\otimes s_2$ and dividing by $t$, we obtain 
\begin{equation}
   \left(\tau_0\sigma_0s_0+ q_x \tau_0\sigma_0s_1-i \tau_0\sigma_3 s_2\right)\chi_\alpha=0,
\end{equation}
which is diagonal in $\tau$ and $\sigma$, but not in $s$. The above equation has non-trivial solutions when the determinant of the term between parenthesis is zero:
\begin{equation}
   \left|\begin{matrix}
       1&q_x-m\\
       q_x+m&1
   \end{matrix}\right|=2-q_x^2=0\rightarrow q_x=\pm \sqrt{2}.
\end{equation}
where $m=\pm 1$ is the eigenstate of $\sigma_3$. Note that $q_x$ does not depend on the eigenvalue $l$ of $\tau_3$. A solution that is localized on the I edge should have negative $q_x$, while one localized on the III edge should have a positive $q_x$. Therefore, for the I edge, $q_x=-\sqrt{2}$, while for the III edge, $q_x=\sqrt{2}$. 

Focusing first on the I edge, we write the localized solutions as $$\chi_{l,m}=\ket{l, m} \otimes\left(a\ket{s_3=1}+b\ket{s_3=-1}\right),$$
such that $a$ and $b$ are given by
\begin{eqnarray}
    \begin{pmatrix}
        1&-\sqrt{2}-m\\
        -\sqrt{2}+m&1
    \end{pmatrix}\begin{pmatrix}
        a\\ b
    \end{pmatrix}=0 \nonumber\\
    \therefore b=\left(m-\sqrt{2}\right)a, \quad a=\sqrt{\frac{1}{4-2m\sqrt{2}}},
\end{eqnarray}
where in the last equality, we used that $a^2+b^2=1$. Next we add the other terms in the Hamiltonian of Eq. \eqref{eq_h_Gamma}, to obtain the energy of the modes localized on the I edge by computing $\chi_{l,m}^\dagger h_{ky} \chi_{l,m}$.  For that, we notice that 
\begin{widetext}
\begin{eqnarray}
     \left[\bra{s_3=1}+\left(m-\sqrt{2}\right)\bra{s_3=-1}\right]&s_1&\left[\ket{s_3=1}+\left(m'-\sqrt{2}\right)\ket{s_3=-1}\right]=m+m'-2\sqrt{2}\\
     \left[\bra{s_3=1}+\left(m-\sqrt{2}\right)\bra{s_3=-1}\right]&s_2&\left[\ket{s_3=1}+\left(m'-\sqrt{2}\right)\ket{s_3=-1}\right]=i(m-m'),
\end{eqnarray}    
\end{widetext}
such that we get
\begin{eqnarray}
  t k_y a\chi_{lm}^\dagger\Gamma_1\chi_{l'm'}&=&- t k_y a\left(\tau_3\right)_{l, l'} \left(\sigma_2\right)_{m, m'}\\
  \Delta\chi_{lm}^\dagger\Gamma_5\chi_{l'm'}&=&0\\
   B_x\chi_{lm}^\dagger\Gamma_6\chi_{l'm'}&=&-\frac{B_x}{\sqrt{2}}\left(\tau_3\right)_{l, l'} \left(\sigma_0\right)_{m, m'},
\end{eqnarray}
where we use that the $\Gamma_6$ term vanishes because it is proportional to $(m-m')\left(\sigma_{0}\right)_{m, m'}=0$. Combining all terms, we arrive at the edge Hamiltonian
\begin{equation}
    h^{\textrm{I}}(k_y)=-t k_y a \tau_3 \otimes \sigma_2-\frac{B_x}{\sqrt{2}}\tau_3\otimes \sigma_0.  
\end{equation}

For the III edge, we instead need to use the solution $q_x=\sqrt{2}$. Performing an equivalent procedure to above, we obtain $$\chi_{lm}=\ket{\tau_3=l}\otimes \ket{\sigma_3=m} \otimes\frac{\ket{s_3=1}-\left(m+\sqrt{2}\right)\ket{s_3=-1}}{\sqrt{4+2m\sqrt{2}}},$$
and 
\begin{equation}
    h^{\textrm{III}}(k_y)=-t k_y a \tau_3 \otimes \sigma_2-\frac{B_x}{\sqrt{2}}\tau_3\otimes \sigma_0.  
\end{equation}

For the II and IV edges, the solutions should instead be localized on $y$. Therefore, we take $h_{0y}=t\Gamma_2+t k_y a \Gamma_1+t \Gamma_4$ to look for localized solutions and consider the perturbation to be $h_{ky}=t k_x a \Gamma_3+\Delta \Gamma_5+B_x\Gamma_6$, making $k_y\rightarrow -i \partial_y$, and look for solutions of 
% \begin{widetext}
    \begin{equation}
    h_{0y}\psi(x, y)=\left[-i t \partial_y a \Gamma_1+t \Gamma_2+t \Gamma_4\right]\psi(x,y)=0.
\end{equation}
Again, we use an ansatz $\varphi_\alpha(y)=\exp(q_y y/a)$ to obtain the matrix equation
\begin{equation}
    \left[q_y \tau_0\sigma_0 s_0-\tau_0\sigma_0s_3+\tau_0\sigma_3s_1\right]\chi_\alpha=0,
\end{equation}
after multiplying $h_{0y}$ by $i/t \, \tau_3 \sigma_2s_1$. In this way, $q_y$ is determined by 
\begin{equation}
    \left|\begin{matrix}q_y-1&m\\m& q_y+1\end{matrix}\right|=q_y^2-2=0\rightarrow q_y=\pm \sqrt{2}. 
\end{equation}

The solution localized on the II edge is the one with $q_y=-\sqrt{2}$, and the one on the IV edge has $q_y=\sqrt{2}$. For the II edge, the solution is given by $$\chi_{l, m}=\ket{l, m} \otimes\frac{\ket{s_3=1}-m\left(1+\sqrt{2}\right)\ket{s_3=-1}}{\sqrt{4+2\sqrt{2}}},$$
and
\begin{equation}
    h^{\textrm{II}}(k_x)=t k_x a \tau_3 \otimes \sigma_2-\frac{1}{\sqrt{2}}B_x\,\tau_3\otimes \sigma_3,  
\end{equation} 
while for the IV edge, the solution is given by $$\chi_{l,m}=\ket{l, m} \otimes\frac{\ket{s_3=1}+m\left(1-\sqrt{2}\right)\ket{s_3=-1}}{\sqrt{4+2\sqrt{2}}},$$
and
\begin{equation}
    h^{\textrm{IV}}(k_x)=t k_x a \tau_3 \otimes \sigma_2-\frac{1}{\sqrt{2}}B_x\,\tau_3\otimes \sigma_3.  
\end{equation}

\bibliography{HOTI}
\end{document}